\documentclass[paper]{JHEP}
\usepackage{epsfig}

\title{On power corrections in the dispersive approach}
\author{G.\ Grunberg\thanks{Research
supported in part by the EC program ``Training and 
Mobility of 
Researchers'', Network ``QCD and Particle Structure'', contract 
ERBFMRXCT980194.}\\
        Centre de Physique Th\'eorique de l' Ecole  
Polytechnique (CNRS UMR C7644),\\
        91128 Palaiseau Cedex, France\\
        E-mail: \email{grunberg@cpht.polytechnique.fr}}
\abstract{Power corrections in QCD (both conventional  and unconventional ones
arising from the ultraviolet region) are discussed within the  infrared 
finite coupling-dispersive approach. It is shown how power corrections 
in Minkowskian quantities can be derived from the corresponding ones in
associated Euclidean quantities through analyticity, allowing a 
parametrization in term of the Euclidean coupling and a renormalon-free
perturbative expansion. It is argued that one should in general expect 
coefficients functions computed in the true non-perturbative vacuum to
differ from the standard perturbative ones, even without assuming new physics.
A phenomenology of   
$1/Q^2$ terms arising from eventual new physics of ultraviolet origin  is
 also set-up. Models for 
non-perturbative contributions to the (universal) QCD coupling are 
suggested. Issues of 
renormalization scheme dependence are  commented upon.}
\preprint{CPTh/S 635.0798}
\begin{document}
\section{Introduction}
The study of power corrections in QCD has been the subject of active
investigations in recent years. Their importance for a precise 
determination of $\alpha_s$ has
been recognized, and  various  techniques (renormalons, finite gluon
 mass, dispersive
approach) have been devised to cope with situations where the 
standard operator product
expansion (OPE) does not apply (for recent reviews see 
ref.\cite{Z1,B1}). In this
paper (which is a revised and extended version of \cite{G1}), I  
investigate 
various issues of the dispersive approach \cite{DMW}, based on the notion
 of 
an infrared (IR)
regular \cite{DW,DKT} universal  QCD coupling $\bar{\alpha}_s$, putting
 emphasis on 
Minkowskian quantities. After a brief
 review of this method in 
section 2, where the possibility of power corrections 
arising from 
non-perturbative  contributions of $\bar{\alpha}_s$ to the
{\em ultraviolet} (UV) region
 is pointed out,
  I first discuss  the ``standard'' power corrections of infrared 
origin, which depend on the (non-perturbative) low energy behavior of
$\bar{\alpha}_s$. It has been recently realized \cite{DLMS,DSW} that a 
renormalon free perturbative expansion can be set up in the general case
 of Minkowskian quantities. I give an explicit check of this statement 
in section 3, where the useful concept of ``IR regularized characteristic
function'' is introduced; the case of a ``causal'' \cite{GGK} perturbative
coupling is also briefly discussed there. In section 4, I give a very 
simple derivation of power corrections to Minkowskian
quantities, as well as of the renormalon-free
perturbative expansion, by relating them through $Q^2$ analyticity to 
the corresponding terms (straightforward to derive) in 
the associated Euclidean  quantities. Section 5 discusses ``ultraviolet''
 power corrections. They may arise either from a
 ${\cal O}(1/k^4)$ power suppressed term in the coupling of ``standard'' 
IR origin, or from more hypothetical (implying new physics) 
${\cal O}(1/k^2)$ terms, which generate
$1/Q^2$  corrections. 	
 I show that a simple phenomenology for  the
channel-dependence of the latter type
of contributions can be set up \cite{G2}. It is also pointed out that 
coefficient 
functions of 
higher dimensional operators computed in the non-perturbative vacuum may differ
from the standard perturbative ones, even without assuming new physics.
In section 6, some  models 
supporting the 
existence of power corrections, not necessarily of the $1/k^2$ type,
 to the running coupling $\bar{\alpha}_s$
itself are presented. A major  difference with  \cite {G1} 
 is that I do not argue
 anymore that $1/Q^2$ terms arise  naturally in the framework of  
\cite {DMW}
from considerations of Landau pole cancellation.  
Section 7 deals with the potentially important renormalization 
sheme (RS) dependence issue, and indicates a possible solution based on 
the
RS independence of the (BLM-like \cite{BLM}) dressed skeleton 
expansion \cite{Lu}. 
A summary and
 conclusions are given in section 8. More technical issues are relegated
to three appendices. A method to derive power corrections to Euclidean
 quantities in the dressed single gluon exchange approximation is 
described in Appendix A. Power corrections to Minkowskian quantities are
discussed in Appendix B. Finally, Appendix C shows how one can express
power corrections to Minkowskian quantities directly in term of the 
Euclidean coupling $\bar{\alpha}_s$ itself.

\section{Parametrization of infrared power corrections}
Consider the contribution to an Euclidean (quark dominated) observable
arising
from dressed virtual
single gluon  exchange, which takes the generic form (after subtraction
of the Born term):
\begin{equation}D(Q^2)\equiv D(\Lambda^2/Q^2)  =\int_{0}^\infty{dk^2
\over k^2}\
\bar{\alpha}_s(\Lambda^2/k^2)\
\Phi_D(k^2/Q^2) \label{eq:D}
\end{equation}
where $\Phi_D$ is the ``distribution function''\cite{N}.
The ``physical'' running coupling $\bar{\alpha}_s(k^2)
\equiv\bar{\alpha}_s(\Lambda^2/k^2)$ is assumed 
to be IR
regular, and thus
must differ from the perturbative coupling  $\bar{\alpha}_s^{PT}(k^2)$ (defined
by the Borel sum eq.(\ref{eq:ab})),
which is assumed\footnote{See
 however the ``causal'' perturbative coupling case in section 3 and the 
remarks in 
section 5.}
 in most of the paper to have a Landau singularity,
 by a non-perturbative
piece $\delta\bar{\alpha}_s(k^2)$ which cancells the singularity:
\begin{equation}\bar{\alpha}_s(k^2)=\bar{\alpha}_s^{PT}(k^2)+
\delta\bar{\alpha}_s(k^2) \label{eq:asplit}
\end{equation}
$\bar{\alpha}_s(k^2)$ should be understood as a universal
 QCD coupling (not to be confused with e.g. the $\overline {MS}$ 
coupling: I use the overbar  to identify this specific 
coupling), an analogue
 of the
Gell-Mann - Low QED effective charge, hopefully defined through an
 extension to QCD of the QED
`` dressed skeleton expansion''\cite{BLM,Lu}. Such a program, which would
 give a firm field
theoretical basis to the ``naive non-abelization'' 
procedure \cite{Broadhurst,BBB} 
familiar in renormalons
calculations, has been initiated in \cite{Watson}. The universal QCD 
coupling
is presently known \cite{CMW,DKT,Watson} only as an expansion in the 
$\overline {MS}$ scheme up to next to leading order\footnote{The 
constant term in the next to leading order expansion of $\bar{\alpha}_s$ 
in the
$\overline{MS}$ scheme  actually differ by a 
factor
of $\pi^2$ in \cite{CMW} and \cite{Watson}. This discrepancy should be 
clarified.}. In the
``large $\beta_0$'' limit of QCD, as implemented through the 
``naive non-abelization'' procedure, $\bar{\alpha}_s(k^2)$  coincides
 with the V-scheme coupling \cite{BLM} (but differs from it at
finite
$\beta_0$).
Consequently:
\begin{eqnarray}D(Q^2)&=&\int_0^\infty{dk^2\over k^2}
\ \bar{\alpha}_s^{PT}(k^2)\
\Phi_D(k^2/Q^2)+
\int_0^\infty{dk^2\over k^2}\ \delta\bar{\alpha}_s(k^2)\
\Phi_D(k^2/Q^2)\nonumber\\
&\equiv&D_{PT}(Q^2)+\delta D(Q^2)
\label{eq:D-split1}\end{eqnarray}
where $\delta D(Q^2)$ contains the power corrections.
To determine the various types of
power contributions, it is appropriate \cite{DW,N} to disentangle long
from short distances \cite{NSVZ,Mueller,Z2} with an IR cut-off  
$\mu_I={\cal O}(\Lambda)$:
\begin{eqnarray}D(Q^2)& =&\int_{0}^{\mu_I^2}{dk^2
\over k^2}\
\bar{\alpha}_s(k^2)\ \Phi_D(k^2/Q^2)
+ \int_{\mu_I^2}^\infty{dk^2\over k^2}\ \bar{\alpha}_s(k^2)\
\Phi_D(k^2/Q^2)\nonumber\\
&\equiv&D_{IR}(Q^2)+D_{UV}(Q^2)
\label{eq:D-split2}\end{eqnarray}
It is convenient to further disentangle perturbative from 
non-perturbative contributions in the short distance part, and set:
\begin{eqnarray}D_{UV}(Q^2)&=&\int_{\mu_I^2}^\infty{dk^2\over k^2}
\ \bar{\alpha}_s^{PT}(k^2)\
\Phi_D(k^2/Q^2)+
\int_{\mu_I^2}^\infty{dk^2\over k^2}\ \delta\bar{\alpha}_s(k^2)\
\Phi_D(k^2/Q^2)\nonumber\\
&\equiv&D_{UV}^{PT}(Q^2)+\delta D_{UV}(Q^2)
\label{eq:Duv-split}\end{eqnarray}
Thus:
\begin{equation}D(Q^2) =D_{IR}(Q^2)+D_{UV}^{PT}(Q^2)+\delta D_{UV}(Q^2)
\label{eq:D-split} \end{equation}
$D_{IR}(Q^2)$ yields, for large $Q^2$, ``long distance ''
power contributions  which correspond to the standard OPE
``condensates''. If the Feynman diagram kernel 
$\Phi_D(k^2/Q^2)$ is 
${\cal O}\left[(k^2/Q^2)^n\right]$ at
small
$k^2$, this piece contributes an  
${\cal O}\left[(\Lambda^2/Q^2)^n\right]$
term from a dimension $n$ condensate, with the normalization given by a
low energy average of the IR regular coupling $\bar{\alpha}_s$
(see Appendix A). 
$D_{UV}^{PT}(Q^2)$
represents a form
of ``regularized perturbation theory '' (choosing the IR cut-off
$\mu_I$ above the  Landau pole), where the long distance part 
  of the perturbative
 contribution has been removed:
\begin{eqnarray}D_{UV}^{PT}(Q^2)&=&\int_{0}^\infty
{dk^2\over k^2}\
\bar{\alpha}_s^{PT}(k^2)\ \Phi_D(k^2/Q^2)-
\int_{0}^{\mu_I^2}{dk^2\over k^2}\
\bar{\alpha}_s^{PT}(k^2)\ \Phi_D(k^2/Q^2)\nonumber\\
&\equiv&D_{PT}(Q^2)-D_{IR}^{PT}(Q^2)
\label{eq:Dreg1}\end{eqnarray}
The last term  $\delta D_{UV}(Q^2)$ in
eq.(\ref{eq:D-split})  yields, unless $\delta\bar{\alpha}_s(k^2)$
 is highly suppressed, additional ``ultraviolet'' power
contributions at large $Q^2$. They are usually neglected, but I shall return to
 them
in section 5 . Note that we have:
\begin{equation}\delta D(Q^2)=\delta D_{IR}(Q^2)+\delta D_{UV}(Q^2)
\label{eq:dDsplit}\end{equation}
with:
\begin{equation}\delta D_{IR}(Q^2)\equiv
\int_0^{\mu_I^2}{dk^2\over k^2}\ \delta\bar{\alpha}_s(k^2)\
\Phi_D(k^2/Q^2)=D_{IR}(Q^2)-D_{IR}^{PT}(Q^2)
\label{eq:dDir}\end{equation}
A derivation of the power corrections in $\delta D(Q^2)$
 can be found in Appendix A.

One would like  to give a parametrization of IR power corrections for 
Minkowskian quantities analoguous to eq.(\ref{eq:D-split}), i.e. in term
 of $\bar{\alpha}_s$ itself. The problem  is that a representation 
such as 
 eq.(\ref{eq:D}) in general does not exist \cite{N}. For a 
sufficiently 
inclusive Minkowskian
quantity $R$, we have instead a representation
 in term of the time like discontinuity of the coupling \cite{BBB,DMW}:
\begin{eqnarray}R(Q^2) &=& \int_0^\infty{d\mu^2
\over\mu^2}\
\bar{\rho}_s(\mu^2)\ \left[{\cal F}_R(\mu^2/Q^2)-
{\cal F}_R(0)\right] \nonumber\\
 &\equiv&\int_0^\infty{d\mu^2\over\mu^2}\ \bar{\alpha}_{eff}(\mu^2)\ 
\dot{{\cal F}}_R(\mu^2/Q^2) \label{eq:R2} 
\end{eqnarray}
where:
\begin{equation}\bar{\rho}_s(\mu^2) = -\frac{1} {2\pi i}
 Disc\{\bar{\alpha}_s(-\mu^2)\}
\equiv -\frac{1}
{2\pi i}\{\bar{\alpha}_s\left[-(\mu^2+i\epsilon)\right]-\bar{\alpha}_s
\left[-(\mu^2-i\epsilon)\right]\} \label{eq:disc}\end{equation}
 is
the time like ``spectral density'', and the``effective 
coupling'' 
$\bar{\alpha}_{eff}(\mu^2)$   is defined  by:
\begin{equation} {d\bar{\alpha}_{eff}\over d\ln \mu^2}=
\bar{\rho}_s(\mu^2)
 \label{eq:eff}\end{equation}
 $\bar{\alpha}_s$ is assumed to satisfy the 
dispersion relation :
\begin{eqnarray}\bar{\alpha}_s(k^2) &=& -\int_0^\infty{d\mu^2\over
\mu^2+k^2}\
\bar{\rho}_s(\mu^2) \nonumber\\
&\equiv&k^2\int_0^\infty{d\mu^2\over(\mu^2+k^2)^2}\ \bar{\alpha}_{eff}
(\mu^2)
\label{eq:disp2}\end{eqnarray}
which implies in particular the absence of Landau singularity.

The ``characteristic 
function''  ${\cal F}_R$ in eq.(\ref{eq:R2})  is computed 
from the  one-loop Feynman diagrams with a finite gluon mass 
$\mu$,
 and $\dot{{\cal F}}_R\equiv - d{\cal F}_R/d\ln \mu^2$. 
It is usually composed of two distinct pieces, for instance:
\begin{equation}{\cal F}_R(\mu^2/Q^2)=
\left\{\begin{array}{lc}
{\cal F}_{R,(-)}({\mu^2/Q^2})&\ (0<\mu^2<Q^2)\\
 {\cal F}_{R,(+)}({\mu^2/Q^2})&\ (\mu^2> Q^2)
\end{array}\right.
\label{eq:char}\end{equation}  
where ${\cal F}_{R,(-)}$ is the sum of a real and a virtual 
contribution,
while 
${\cal F}_{R,(+)}$ contains only
the virtual contribution, and may vanish identically, as in the case of 
thrust.
 This feature prevents (see the comment below eq.(\ref{eq:FDdisp})) a 
representation
 of $R$ 
similar to
eq.(\ref{eq:D}) to be 
reconstructed from eq.(\ref{eq:R2}) using analyticity. 
Nevertheless, 
as pointed out in \cite{DLMS,DSW},
it is still possible to
parametrize the IR power corrections in term of $\bar{\alpha}_s$ (a
parametrization of power corrections in term of quantities related to
the ``Minkowskian coupling'' $\bar{\alpha}_{eff}$ is possible, but 
cumbersome (see Appendix B)). 

The first 
observation \cite{G1} is that the power correction piece $\delta R(Q^2)$
 in: 
\begin{equation}R(Q^2)= R_{PT}(Q^2)+\delta R(Q^2)
\label{eq:Rsplit1}\end{equation} 
(where $R_{PT}(Q^2)$ is defined to be the Borel sum, 
see eq.(\ref{eq:Rb}) below)
can be expressed as an integral over the non-perturbative modification 
$\delta\bar{\alpha}_s$ of the coupling in eq.(\ref{eq:asplit}).
Provided 
$\delta\bar{\alpha}_s(k^2)$ decreases sufficiently fast at large $k^2$, 
one can indeed show (Appendix C) that:
\begin{equation}\delta R(Q^2)\simeq \int_0^\infty{dk^2\over k^2}\ 
\delta\bar{\alpha}_s(k^2)\
\Phi_R(k^2/Q^2) \label{eq:dR}\end{equation}
where $\Phi_R(k^2/Q^2)$ is the discontinuity  at $\mu^2=-k^2<0$
of the ``low gluon mass'' piece
${\cal F}_{R,(-)}$ of the characteristic function.
The proof of eq.(\ref{eq:dR}) is actually not straightforward, if one assumes
 the perturbative coupling
$\bar{\alpha}_s^{PT}$  (hence also $\delta\bar{\alpha}_s$) has a Landau
singularity. The reason is that, if $\bar{\alpha}_s^{PT}$ is  not ``causal'',
i.e. does not satisfy eq.(\ref{eq:disp2}) with
 $\bar{\alpha}_{eff}\rightarrow \bar{\alpha}_{eff}^{PT}$,
  $\delta\bar{\alpha}_s$ is
 {\em  not\/} related to its time-like analogue 
$\delta\bar{\alpha}_{eff}$ (where $\bar{\alpha}_{eff}=
\bar{\alpha}_{eff}^{PT}+
\delta\bar{\alpha}_{eff}$) by the
dispersion relation eq.(\ref{eq:disp2}). Moreover, $R_{PT}$ and
 $\delta R$ are {\em not} given (see Appendix
 B)  by eq.(\ref{eq:R2}) with
$\bar{\alpha}_{eff}$ substituted respectively by 
$\bar{\alpha}_{eff}^{PT}$ and $\delta\bar{\alpha}_{eff}$.

In a second step, one introduces an IR
cutoff $\mu_I$ into eq.(\ref{eq:dR}) and substitute 
$\delta\bar{\alpha}_s(k^2)=\bar{\alpha}_s(k^2)-
\bar{\alpha}_s^{PT}(k^2)$ in the low energy piece. Thus,
neglecting the high energy integral above $\mu_I^2$, one gets:
\begin{equation}\delta R(Q^2)\simeq \delta R_{IR}(Q^2)
\label{eq:dR1}\end{equation}
with:
\begin{equation}\delta R_{IR}(Q^2)\equiv \int_0^{\mu_I^2}
{dk^2\over k^2}\ \delta\bar{\alpha}_s(k^2)\
\Phi_R(k^2/Q^2)=R_{IR}(Q^2)-R_{IR}^{PT}(Q^2)
\label{eq:dRir}\end{equation}
where:
\begin{equation}R_{IR}(Q^2)\equiv \int_0^{\mu_I^2}{dk^2\over k^2}
\ \bar{\alpha}_s(k^2)\ \Phi_R(k^2/Q^2)
\label{eq:Rir}\end{equation}
and:
\begin{equation}R_{IR}^{PT}(Q^2)\equiv \int_0^{\mu_I^2}{dk^2\over k^2}
\ \bar{\alpha}_s^{PT}(k^2)\ \Phi_R(k^2/Q^2)
\label{eq:Rirpt}\end{equation}

One deduces \cite{DLMS,DSW}:
\begin{eqnarray} R(Q^2)&\simeq&R_{PT}(Q^2)+\delta R_{IR}(Q^2)\nonumber\\
&=&R_{IR}(Q^2)
+R_{UV}^{PT}(Q^2)\label{eq:R-split}\end{eqnarray}
with:
\begin{equation}R_{UV}^{PT}(Q^2)\equiv
R_{PT}(Q^2)-R_{IR}^{PT}(Q^2)\label{eq:Rreg1}\end{equation}
Eq.(\ref{eq:R-split}), which is correct only if 
$\delta\bar{\alpha}_s(k^2)$ decreases fast enough at large $k^2$,
 is the analogue of eq.(\ref{eq:D-split}) (with the last ``UV'' piece
neglected), and the
``infrared'' power corrections are again parametrized in term of a low
energy average of $\bar{\alpha}_s$. The only difference is that the 
``regularized perturbation theory'' piece $R_{UV}^{PT}$
(just as $R_{PT}$)  can no more be
written as an integral, cut-off in the infrared, over the perturbative 
part of the coupling, although it is still a renormalon free quantity.

\section{Cancellation of IR renormalons} 
To check the latter property explicitly, it is   convenient to
revert to the original representation eq.(\ref{eq:R2}), and express the
 Borel transform of 
$R_{UV}^{PT}$ in term of that of 
$\bar{\alpha}_{eff}^{PT}$. 
One  introduces the (RS invariant \cite{G3,B2,BY}) Borel
 representations\footnote{An {\em exact} expression 
for $\tilde{\alpha}_s(z)$ in the case $\bar{\alpha}_s^{PT}(k^2)$ 
satisfies the two loop renormalization group equation is given
in \cite{G3}.}
of $\bar{\alpha}_s^{PT}(k^2)$ and $\bar{\alpha}_{eff}^{PT}(\mu^2)$:
\begin{equation}\bar{\alpha}_s^{PT}(k^2)=\int_0^\infty dz 
\exp\left(-z\beta_0\ln{k^2 \over\Lambda^2}\right)\
 \tilde{\alpha}_s(z) \label{eq:ab}\end{equation}
and:
\begin{equation}\bar{\alpha}_{eff}^{PT}(\mu^2)=
\int_0^\infty dz
\exp\left(-z\beta_0\ln{\mu^2 \over\Lambda^2}\right)\
 \tilde{\alpha}_{eff}(z)\label{eq:aeffb}\end{equation}
($\beta_0$ is (minus) the one  loop beta function coefficient)
where the (RS invariant) Borel transforms
$\tilde{\alpha}_{eff}(z)$ and $\tilde{\alpha}_s(z)$ are related
by \cite{G1,BY} :
\begin{equation}\tilde{\alpha}_{eff}(z)={\sin(\pi\beta_0 z)
\over\pi\beta_0 z}\ \tilde{\alpha}_s(z)\label{eq:b-b}\end{equation}
Substituting 
$\bar{\alpha}_{eff}(\mu^2)$ in eq.(\ref{eq:R2}) with 
$\bar{\alpha}_{eff}^{PT}(\mu^2)$ as 
given 
by eq.(\ref{eq:aeffb}), one gets the (RS invariant) Borel representation:
\begin{equation}R_{PT}(Q^2)=\int_0^\infty 
dz\ \tilde{\alpha}_{eff}(z)\ 
\left[\int_0^{\infty}{d\mu^2\over \mu^2}\
\dot{{\cal F}}_R(\mu^2/Q^2)\ 
\exp\left(-z\beta_0\ln{\mu^2\over
\Lambda^2}\right)\right]\label{eq:Rb}\end{equation}
Note that, since the representation eq.(\ref{eq:aeffb}) is valid only for 
$\mu^2>\Lambda^2$, and $\bar{\alpha}_{eff}^{PT}(\mu^2)$ has a non-trivial
IR fixed point \cite{BBB},
the Borel sum $R_{PT}(Q^2)$ is  {\em different} \cite{G4,DU}, 
as mentionned below eq.(\ref{eq:dR}),
from the corresponding ``gluon mass'' integral \cite{BBB}:
\begin{equation}R_{APT}(Q^2)\equiv
\int_0^\infty{d\mu^2\over\mu^2}
\ \bar{\alpha}_{eff}^{PT}(\mu^2)\ \dot{{\cal F}}_R(\mu^2/Q^2)
\label{eq:Rapt}\end{equation} 
Next consider $R_{IR}(Q^2)$ (eq.(\ref{eq:Rir})).
Using the dispersion relation eq.(\ref{eq:disp2}),
 one gets the ``Minkowskian'' representation 
(analogue
of eq.(\ref{eq:R2})):
\begin{equation}R_{IR}(Q^2)=\int_0^\infty{d\mu^2\over\mu^2}
\ \bar{\alpha}_{eff}(\mu^2)\ 
\dot{{\cal F}}_{R,IR}(\mu^2,Q^2) \label{eq:Rir1}
\end{equation}
with:
\begin{equation}\dot{{\cal F}}_{R,IR}(\mu^2,Q^2)\equiv
\mu^2\int_0^{\mu_I^2}{dk^2\over(k^2+\mu^2)^2}\ 
\Phi_R(k^2/Q^2)\label{eq:Fir}\end{equation}
and $\dot{{\cal F}}_{R,IR}(\mu^2,Q^2)$ is written as a function of two
variables to emphasize that  a third scale ($\mu_I$) is involved.
Substituting $\bar{\alpha}_{eff}$ in eq.(\ref{eq:Rir1})
with eq.(\ref{eq:aeffb}), 
one  finds similarly the Borel representation of $R_{IR}^{PT}(Q^2)$:
\begin{equation}R_{IR}^{PT}(Q^2)=\int_0^\infty 
dz\ \tilde{\alpha}_{eff}(z)\ 
\left[\int_0^{\infty}{d\mu^2\over \mu^2}\
\dot{{\cal F}}_{R,IR}(\mu^2,Q^2)\ \exp\left(-z\beta_0
\ln{\mu^2\over
\Lambda^2}\right)\right]\label{eq:Rirb}\end{equation}
(using the original definition of $R_{IR}^{PT}$ in eq.(\ref{eq:Rirpt}),
 one could have also written the Borel transform in term of 
$\Phi_R$, but this is less useful for Minkowskian quantities).
One deduces from eq.(\ref{eq:Rreg1}):
\begin{equation}R_{UV}^{PT}(Q^2)=\int_0^\infty 
dz\ \tilde{\alpha}_{eff}(z)\ 
\left[\int_0^{\infty}{d\mu^2\over \mu^2}\
\dot{{\cal F}}_{R,UV}(\mu^2,Q^2)\ \exp
\left(-z\beta_0\ln{\mu^2\over
\Lambda^2}\right)\right]\label{eq:Rregb}\end{equation}
where:
\begin{equation}\dot{{\cal F}}_{R,UV}(\mu^2,Q^2)
\equiv
\dot{{\cal F}}_R(\mu^2/Q^2)-
\dot{{\cal F}}_{R,IR}(\mu^2,Q^2)\label{eq:Freg}
\end{equation}
is the ``IR regularized'' characteristic function. Again, the Borel sums
$R_{IR}^{PT}$ and $R_{UV}^{PT}$ are  different from
the corresponding ``gluon mass'' integrals 
\begin{equation}R_{IR}^{APT}(Q^2)\equiv
\int_0^\infty{d\mu^2\over\mu^2}
\ \bar{\alpha}_{eff}^{PT}(\mu^2)\ 
\dot{{\cal F}}_{R,IR}(\mu^2, Q^2)
\label{eq:Raptir}\end{equation}
 and
\begin{equation}R_{UV}^{APT}(Q^2)\equiv
\int_0^\infty{d\mu^2\over\mu^2}
\ \bar{\alpha}_{eff}^{PT}(\mu^2)\ \dot{{\cal F}}_{R,UV}(\mu^2, Q^2)
\label{eq:Raptuv}\end{equation}
The effect of the subtracted term in eq.(\ref{eq:Freg}) is to remove
any potential non-analytic term in the small $\mu^2$ expansion of 
$\dot{{\cal F}}_{R,UV}(\mu^2,Q^2)$ (now distinct  from its large $Q^2$
behavior, since there is a 
third scale $\mu_I$), by introducing an IR cutoff $\mu_I$
 in the dispersive integral
over $\Phi_R$ (see eq.(\ref{eq:Fdisp1}) or (\ref{eq:Fdisp2}), and
   also eq.(\ref{eq:FDreg}) below). It follows 
that the renormalons singularities in
eq.(\ref{eq:Rregb}) can only be simple poles, which are however cancelled
by the zeroes of the sin factor in eq.(\ref{eq:b-b})
($\tilde{\alpha}_s(z)$ itself is assumed throughout to have no 
renormalons). The Borel sum 
eq.(\ref{eq:Rregb}) is thus renormalon-free and unambiguous, as expected.

A similar formalism also applies \cite{BBB,DMW}
 to Euclidean 
quantities. Indeed, proceeding as for eq.(\ref{eq:Rir1}) and 
using the dispersion relation eq.(\ref{eq:disp2}) into
the ``Euclidean'' representation eq.(\ref{eq:D}) gives the
``Minkowskian'' representation:
\begin{equation}D(Q^2)=\int_0^\infty{d\mu^2\over\mu^2}\ 
\bar{\alpha}_{eff}(\mu^2)\ 
\dot{{\cal F}}_D(\mu^2/Q^2) \label{eq:D1}\end{equation}
where the Euclidean characteristic function ${\cal F}_D$ is
related to the corresponding distribution function $\Phi_D$
by the dispersion relation \cite{BBB1,N}:
\begin{equation}\dot{{\cal F}}_D(\mu^2/Q^2)=
\mu^2\int_0^\infty{dk^2\over(k^2+\mu^2)^2}\ 
\Phi_D(k^2/Q^2)\label{eq:FDdisp}\end{equation}
Note that ${\cal F}_D$, at the difference of ${\cal F}_R$, must be made 
of
a single piece, which shows that an Euclidean representation is indeed
 not possible for $R$.
We then have:
\begin{equation}D_{UV}^{PT}(Q^2)=
\int_0^\infty 
dz\ \tilde{\alpha}_{eff}(z)\ 
\left[\int_0^{\infty}{d\mu^2\over \mu^2}\
\dot{{\cal F}}_{D,UV}(\mu^2, Q^2)
\ \exp\left(-z\beta_0\ln{\mu^2\over
\Lambda^2}\right)\right]\label{eq:Dregb}\end{equation}
with:
\begin{equation}\dot{{\cal F}}_{D,UV}(\mu^2, Q^2)\equiv
\mu^2\int_{\mu^2_I}^\infty{dk^2\over(k^2+\mu^2)^2}\ 
\Phi_D(k^2/Q^2)\label{eq:FDreg}\end{equation}
where the IR cut-off is explicit. It is also clear that the small $\mu^2$
behavior of $\dot{{\cal F}}_{D,UV}(\mu^2, Q^2)$ is
analytic. The representation eq.(\ref{eq:Dregb}) shall be used in 
section 4.

\noindent\underline{Case of a ``causal'' perturbative coupling}:
one can show that eq.(\ref{eq:R-split}) is still valid, and is in fact 
{\em exact}
\footnote{For Euclidean quantities, this statement follows 
from the results in \cite{G4,DU}.}
(with $R_{UV}^{PT}(Q^2)$ as in eq.(\ref{eq:Rregb})),  if 
$\delta\bar{\alpha}_s(k^2)\equiv 0$. Then $\bar{\alpha}_s^{PT}$ is
``causal'' and
satisfies by itself the dispersion relation eq.(\ref{eq:disp2}). This is known
to occur in QCD for a large enough number of flavors, where the perturbative
coupling has a non-trivial IR fixed point and no (real or 
complex) Landau singularity \cite{GGK}.
 In such a case,
$\bar{\alpha}_s^{PT}=\bar{\alpha}_s^{APT}$, where 
$\bar{\alpha}_s^{APT}$ is the ``analytic'' perturbation theory coupling
\cite{SS,G1,GGK,BBB} of eq.(\ref{eq:areg}) and (\ref{eq:APT}), and we have:
\begin{equation}R(Q^2)=R_{APT}(Q^2)=R_{IR}^{APT}(Q^2)+R_{UV}^{APT}(Q^2)
\label{eq:Rapt-split}\end{equation}
where $R_{APT}$, $R_{IR}^{APT}$ and $R_{UV}^{APT}$ are the ``gluon mass''
integrals of
eq.(\ref{eq:Rapt}),
(\ref{eq:Raptir}) and (\ref{eq:Raptuv}).
But, using
eq.(\ref{eq:areg}), we get:
\begin{equation}R_{IR}^{APT}(Q^2)= 
\int_0^{\mu_I^2}{dk^2\over k^2}
\ \bar{\alpha}_s^{APT}(k^2)\ \Phi_R(k^2/Q^2)=
\int_0^{\mu_I^2}{dk^2\over k^2}
\ \bar{\alpha}_s^{PT}(k^2)\ \Phi_R(k^2/Q^2)
\label{eq:Rapt-ir}\end{equation}
where in the second step I used that $\bar{\alpha}_s^{PT}$ is causal.
On the other hand one can show, even if  $\bar{\alpha}_s^{PT}$
is not causal (see Appendix B) that power corrections in 
$R_{UV}^{APT}$ involve only {\em analytic} moments of
$\bar{\alpha}_{eff}^{PT}$ (the $b_n^{APT}$'s of eq.(\ref{eq:b_{PT}})),
 as a 
consequence
of the analytic behavior of $\dot{{\cal F}}_{R,UV}(\mu^2,Q^2)$ at small
$\mu^2$ (and despite the
 presence of non-analytic terms in the large $Q^2$ behavior). These moments
 turn
 out to vanish if $\bar{\alpha}_s^{PT}$ is causal.
Thus for a causal coupling
$R_{UV}^{APT}(Q^2)$ (eq.(\ref{eq:Raptuv})) coincides
with the Borel sum $R_{UV}^{PT}(Q^2)$  (eq.(\ref{eq:Rregb})), 
and consequently:
\begin{equation}R_{APT}(Q^2)=R_{IR}^{APT}(Q^2)+R_{UV}^{PT}(Q^2)
\label{eq:Rapt-split1}\end{equation}
which is just eq.(\ref{eq:R-split}) in this case.
One can similarly show, if the perturbative coupling is ``causal'',
that analytic
terms in the small $\mu^2$ behavior of 
$\dot{{\cal F}}_R(\mu^2/Q^2)$ can contribute no power
 corrections, and therefore  
$R_{APT}(Q^2)$ differs from the Borel sum $R_{PT}(Q^2)$  only by the
 ``OPE-compatible'' power corrections arising from the non-analytic 
terms. Note also $R_{IR}^{PT}(Q^2)$ is 
no more identical, due to the
non-trivial IR fixed point of $\bar{\alpha}_s^{PT}$, to the right
hand side of eq.(\ref{eq:Rapt-ir}),
 and should rather be replaced  in eq.(\ref{eq:Rirpt}) and 
(\ref{eq:Rreg1}) by
the corresponding Borel sum 
eq.(\ref{eq:Rirb}).

\section{The Euclidean-Minkowskian connection}

The regularization procedure 
described in the
previous section is rather formal, and provides no physical 
picture of $\dot{{\cal F}}_{R,UV}$ as an IR cutoff Feynman diagram 
(as opposed to $\dot{{\cal F}}_{D,UV}$).
A more transparent interpretation can be given if the Minkowskian 
quantity $R$ is 
related to the time-like discontinuity of an 
Euclidean quantity $D$, i.e. I shall assume $D$ satisfies the dispersion
relation:
\begin{equation}D(Q^2)=Q^2\int_0^\infty{dQ'^2\over(Q'^2+Q^2)^2}\ R(Q'^2)
\label{eq:D-R}\end{equation}
which implies (if $R(Q^2)$ vanishes  at $Q^2=0$ ) 
the inverse relation:
\begin{equation}R(Q^2)={1\over 2\pi i} \oint_{|Q'^2|=Q^2}
{dQ'^2\over Q'^2}\
D(Q'^2)\label{eq:R-D}\end{equation}
The main point of this section is that each term in 
eq.(\ref{eq:R-split}) can be obtained from 
the corresponding ones in eq.(\ref{eq:D-split}) through the relation 
eq.(\ref{eq:R-D}) (for large enough $Q^2$), i.e. by taking their
(integrated) time-like discontinuity, after analytic continuation 
to complex $Q^2$ (the latter formulation being valid for all $Q^2$).
Note that these statements do not imply that these terms are
necessarily related by the dispersion relation eq.(\ref{eq:D-R}).

I first observe that eq.(\ref{eq:D1}) and (\ref{eq:FDdisp}) allow an 
analytic continuation\footnote{Alternatively, one could continue 
$D(Q^2)$
 to complex 
$\Lambda^2$ using eq.(\ref{eq:D}), keeping $k^2$ and $Q^2$ real
(this is a simplified version of the method of \cite{N}).
Similarly, considering the analytic properties of $D(Q^2)$ in 
eq.(\ref{eq:D-R}) with respect to $\Lambda^2$, i.e. essentially with 
respect to the variable $1/Q^2$, one can derive eq.(\ref{eq:R-D}) from 
the alternative assumption (which involves no non-perturbative physics)
that $R(Q^2)$ vanishes for $Q^2\rightarrow\infty$.}
of $D(Q^2)$  
to complex $Q^2$
(through continuation
 of the 
integrand, keeping $\mu^2$ (and $\Lambda^2$) real),
 and imply that $D(Q^2)$ does
satisfy the dispersion relation eq.(\ref{eq:D-R}). Indeed, performing
the change of variable: $k^2={\mu^2 Q^2\over Q'^2}$,
eq.(\ref{eq:FDdisp}) can also
be written  as:
\begin{equation}\dot{{\cal F}}_D(\mu^2/Q^2)=
Q^2\int_0^\infty{dQ'^2\over(Q'^2+Q^2)^2}\
 \Phi_D(\mu^2/Q'^2)\label{eq:FDdisp1}\end{equation}
Inserting eq.(\ref{eq:FDdisp1}) into eq.(\ref{eq:D1}) reproduces
eq.(\ref{eq:D-R}), with:
\begin{equation}R(Q^2)=\int_0^\infty{d\mu^2\over\mu^2}\ 
\bar{\alpha}_{eff}(\mu^2)\ 
\Phi_D(\mu^2/Q^2)\label{eq:R1}\end{equation}
As a byproduct, we also learn that \cite{N}:
\begin{equation}\Phi_D(\mu^2/Q^2)\equiv
\dot{{\cal F}}_R(\mu^2/Q^2)\label{eq:ker}\end{equation}
Furthermore, the inverse relation eq.(\ref{eq:R-D}) is also
satisfied, since $R(Q^2)$ vanishes for $Q^2\rightarrow\infty$ 
(see previous
footnote). Alternatively, I note that eq.(\ref{eq:FDdisp})
 implies, since $\Phi_D(\mu^2/Q^2)$ vanishes at 
$\mu^2=0$:
\begin{equation}\Phi_D(\mu^2/Q^2)=
{1\over 2\pi i} \oint_{|k^2|=\mu^2}
{dk^2\over k^2}\
\dot{{\cal F}}_D(k^2/Q^2)
\label{eq:Dker1}\end{equation}
which can be written with the previous change of variable:
\begin{equation}\dot{{\cal F}}_R(\mu^2/Q^2)=
{1\over 2\pi i} \oint_{|Q'^2|=Q^2}
{dQ'^2\over Q'^2}\
\dot{{\cal F}}_D(\mu^2/Q'^2)
\label{eq:Dker}\end{equation}
From eq.(\ref{eq:D1}) and
(\ref{eq:Dker})
  one then deduces:
\begin{eqnarray}{1\over 2\pi i} \oint_{|Q'^2|=Q^2}
{dQ'^2\over Q'^2}\
D(Q'^2)&=&\int_0^\infty{d\mu^2\over\mu^2}\ 
\bar{\alpha}_{eff}(\mu^2)\ 
{1\over 2\pi i} \oint_{|Q'^2|=Q^2}
{dQ'^2\over Q'^2}\
\dot{{\cal F}}_D(\mu^2/Q'^2)\nonumber\\
&=&\int_0^\infty{d\mu^2\over\mu^2}\ 
\bar{\alpha}_{eff}(\mu^2)\ 
\dot{{\cal F}}_R(\mu^2/Q^2)=R(Q^2)\label{eq:R-D1}
\end{eqnarray}
which proves eq.(\ref{eq:R-D}).

Consider next the term  $D_{IR}(Q^2)$ in eq.(\ref{eq:D-split}),
 which contains
 the ``IR'' power corrections.
For $\mu_I^2<Q^2$, one can identify $\Phi_D(k^2/Q^2)$
with  
the ``low gluon mass piece'' of
$\dot{{\cal F}}_R(k^2/Q^2)$, and 
$D_{IR}(Q^2)$ (eq.(\ref{eq:D-split2})) may be analytically continued to
 complex $Q^2$ 
through the continuation of the integrand,
 keeping $k^2$ 
(and $\Lambda^2$) real. Thus,
for $Q^2>\mu_I^2$:
\begin{equation}{1\over 2\pi i} \oint_{|Q'^2|=Q^2}
{dQ'^2\over Q'^2}\
D_{IR}(Q'^2)
=\int_0^{\mu_I^2}{dk^2\over k^2}
\ \bar{\alpha}_s(k^2)\ {1\over 2\pi i} \oint_{|Q'^2|=Q^2}
{dQ'^2\over Q'^2}\ \Phi_D(k^2/Q'^2)
\label{eq:Rir2}\end{equation}

Now, performing the change of   variable 
$Q'^2={k^2 Q^2\over k'^2}$, we get:
\begin{equation}{1\over 2\pi i}\oint_{|Q'^2|=Q^2}
{dQ'^2\over Q'^2}\ \Phi_D(k^2/Q'^2)={1\over 2\pi i}
\oint_{|k'^2|=k^2}
{dk'^2\over k'^2}\ \dot{{\cal F}}_R(k'^2/Q^2)
\label{eq:kern}\end{equation}
where I also used eq.(\ref{eq:ker}).
Furthermore, assuming the absence of {\em complex}
 singularities in 
$\dot{{\cal F}}_R(k^2/Q^2)$ for small 
enough ``gluon mass'' $|k^2|$ (which is certainly the case if the 
low $k^2$ expansion of $\dot{{\cal F}}_R(k^2/Q^2)$ involves only
elementary functions), 
 Cauchy theorem yields for small enough $k^2$ ($\Phi_R$ vanishes
at $k^2=0$):
\begin{equation}
{1\over 2\pi i} \oint_{|k'^2|=k^2}
{dk'^2\over k'^2}\ \dot{{\cal F}}_R(k'^2/Q^2)
=\Phi_R(k^2/Q^2)\label{eq:Rker}\end{equation}
Eq.(\ref{eq:Rker}) is the analogue of eq.(\ref{eq:Dker1}); being
a finite ``gluon mass'' sum rule, it does not depend on the high $k^2$
 behavior of $\dot{{\cal F}}_R$, and in particular still holds
in  presence 
of subtractions. We thus obtain, for large enough $Q^2$:
\begin{equation}\Phi_R(k^2/Q^2)=
{1\over 2\pi i} \oint_{|Q'^2|=Q^2}
{dQ'^2\over Q'^2}\ \Phi_D(k^2/Q'^2)
\label{eq:Rker1}\end{equation}
which shows (see eq.(\ref{eq:Rir})) that we have:
\begin{equation}R_{IR}(Q^2)={1\over 2\pi i} \oint_{|Q'^2|=Q^2}
{dQ'^2\over Q'^2}\
D_{IR}(Q'^2)\label{eq:(R-D)ir}\end{equation}
Note that if 
$\Phi_D$ were
made of a single piece at {\em all} scales, one could apply the same 
argument 
to the {\em whole\/} $D(Q^2)$ to get an expression for $R$ in term of 
$\bar{\alpha}_s$.

Moreover, one can also  show that:
\begin{equation}R_{UV}^{PT}(Q^2)={1\over 2\pi i} \oint_{|Q'^2|=Q^2}
{dQ'^2\over Q'^2}\
D_{UV}^{PT}(Q'^2)\label{eq:R-Dreg}\end{equation}
The analytic continuation of $D_{UV}^{PT}(Q^2)$ is now more tricky, since 
one has to treat differently the two terms on the right hand side of 
eq.(\ref{eq:Dreg1}), namely (since $\Phi_D$ is discontinuous) 
continue $D_{PT}(Q^2)$ with respect to 
$\Lambda^2$ (see previous footnote), and $D_{IR}^{PT}(Q^2)$ with respect
 to $Q^2$ (this is only
heuristic, since both terms are separately not well defined).   
Indeed, the operator 
${1\over 2\pi i} \oint_{|Q'^2|=Q^2}{dQ'^2\over Q'^2}$
 applied to $D_{IR}^{PT}(Q^2)$
 formally 
reproduces $R_{IR}^{PT}(Q^2)$, as suggested by the previous result.
 On the other hand, applying this operator
 to the formal 
(RS invariant) Borel representation:
\begin{equation}D_{PT}(Q^2)=\int_0^\infty dz \exp\left(-z\beta_0
\ln{Q^2 \over\Lambda^2}\right)\ \tilde{D}(z)\label{eq:Dbor}\end{equation}
yields quite generally:
\begin{equation}R_{PT}(Q^2)=\int_0^\infty dz \exp\left(-z\beta_0
\ln{Q^2 \over\Lambda^2}\right)\ \tilde{R}(z)\label{eq:Rbor}\end{equation}
with:
\begin{equation}\tilde{R}(z)= {\sin(\pi\beta_0 z)\over\pi\beta_0 z }
\ \tilde{D}(z)\label{eq:RDbor}\end{equation}
which is the correct  expected  relation \cite{BY}.

A clearer derivation is afforded by 
using eq.(\ref{eq:Dregb}) and observing that 
$\dot{{\cal F}}_{D,UV}(\mu^2,Q^2)$, analytically continued to complex
$Q^2$, satisfies the relation:
\begin{equation}{1\over 2\pi i} \oint_{|Q'^2|=Q^2}{dQ'^2\over Q'^2}\ 
\dot{{\cal F}}_{D,UV}(\mu^2,Q'^2)=
\dot{{\cal F}}_{R,UV}(\mu^2,Q^2)
\label{eq:F-DR}\end{equation}
Eq.(\ref{eq:R-Dreg}) then follows by applying the operator
${1\over 2\pi i} \oint_{|Q'^2|=Q^2}{dQ'^2\over Q'^2}$ to  
eq.(\ref{eq:Dregb}) and comparing with eq.(\ref{eq:Rregb}).
To check eq.(\ref{eq:F-DR}), I note that:
\begin{equation}\dot{{\cal F}}_{D,UV}(\mu^2,Q^2)=
\dot{{\cal F}}_D(\mu^2/Q^2)-\dot{{\cal F}}_{D,IR}(\mu^2,Q^2)
\label{eq:F-Dsplit}\end{equation}
where:
\begin{equation}\dot{{\cal F}}_{D,IR}(\mu^2,Q^2)\equiv
\mu^2\int_0^{\mu_I^2}{dk^2\over(k^2+\mu^2)^2}\ 
\Phi_D(k^2/Q^2)\label{eq:F-Dir}\end{equation}
Furthermore $\Phi_D$, hence also $\dot{{\cal F}}_{D,IR}$,
can be analytically continued to complex $Q^2$ for large enough
 $Q^2/\mu_I^2$ (keeping $k^2$ and $\mu^2$ real), and we have:
\begin{eqnarray}{1\over 2\pi i} \oint_{|Q'^2|=Q^2}{dQ'^2\over Q'^2}\ 
\dot{{\cal F}}_{D,IR}(\mu^2,Q'^2)&=&\mu^2\int_0^{\mu_I^2}
{dk^2\over(k^2+\mu^2)^2}\ 
\oint_{|Q'^2|=Q^2}{dQ'^2\over Q'^2}\ 
\Phi_D(k^2/Q'^2)\nonumber\\
&=&\mu^2\int_0^{\mu_I^2}{dk^2\over(k^2+\mu^2)^2}\
 \Phi_R(k^2/Q^2)\nonumber\end{eqnarray}
(where I used eq.(\ref{eq:Rker1})), i.e.:
\begin{equation}{1\over 2\pi i} \oint_{|Q'^2|=Q^2}{dQ'^2\over Q'^2}\ 
\dot{{\cal F}}_{D,IR}(\mu^2,Q'^2)
=\dot{{\cal F}}_{R,IR}(\mu^2,Q^2)
\label{eq:F-DRir}\end{equation}
which, together with eq.(\ref{eq:Dker}) and eq.(\ref{eq:F-Dsplit}),
and comparing with eq.(\ref{eq:Freg}), proves eq.(\ref{eq:F-DR}).
Note that eq.(\ref{eq:F-DRir}) could have been used to derive
eq.(\ref{eq:(R-D)ir}) starting from the ``Minkowskian representation'':
\begin{equation}D_{IR}(Q^2)=\int_0^\infty{d\mu^2\over\mu^2}
\ \bar{\alpha}_{eff}(\mu^2)\ 
\dot{{\cal F}}_{D,IR}(\mu^2,Q^2) \label{eq:Dir1}
\end{equation}
and comparing with eq.(\ref{eq:Rir1}).
 
Finally, 
a similar 
relation holds between $\delta D_{UV}$ in eq.(\ref{eq:Duv-split})
 and  the ``UV'' piece $\delta R_{UV}$ defined by:
\begin{equation}\delta R(Q^2)\equiv\delta R_{IR}(Q^2)+
\delta R_{UV}(Q^2)\label{eq:dR3}\end{equation}
or, equivalently:
\begin{equation}R(Q^2)\equiv R_{IR}(Q^2)+R_{UV}^{PT}(Q^2)+
\delta R_{UV}(Q^2)\label{eq:R-split1}\end{equation}
As a consequence of eq.(\ref{eq:R-D}), (\ref{eq:(R-D)ir}) and
(\ref{eq:R-Dreg}) one indeed gets:
\begin{equation}\delta R_{UV}(Q^2)={1\over 2\pi i} 
\oint_{|Q'^2|=Q^2}{dQ'^2\over Q'^2}\
\delta D_{UV}(Q'^2)\label{eq:dR-dDreg}\end{equation}
However, at the difference of $\delta R_{IR}$ (eq.(\ref{eq:dRir})),
 $\delta R_{UV}$ (and
$\delta R$) cannot in general be expressed as integrals over 
$\delta\bar{\alpha}_s$ (an exception is 
$\delta R_{APT}(Q^2)$, see Appendix C). 
This relation will be used in the next section.

On a more formal level, I note that the present method
may be applied to any Minkowskian quantity $R(Q^2)$ (even not 
related 
to the discontinuity of a genuine Euclidean correlation function), since
one can always associate to any given $R(Q^2)$ a corresponding
``Euclidean'' $D(Q^2)$
{\em defined} by eq.(\ref{eq:D-R}) and (\ref{eq:ker}). This method also
shows that the dispersive approach, applied to Minkowskian quantities,
can be viewed as an extension of the quark-hadron duality as implemented
through finite energy sum rules.

\section{Ultraviolet  power corrections}
I now turn to the third, ``ultraviolet'' contributions in 
eq.(\ref{eq:D-split}) and (\ref{eq:R-split1}). These pieces are usually
neglected,
 on the ground they may yield power contributions unrelated to the OPE 
condensates,
 and consequently it
is often assumed that $\delta\bar{\alpha}_s(k^2)$ is very highly 
suppressed in 
the UV region. However, as argued in ref.\cite{G1,AZ}, the opposite 
assumption violates no known principle, and  (moderately suppressed) 
power 
corrections to $\bar{\alpha}_s(k^2)$ are actually quite naturally 
expected
(see section 6). I will discuss two illustrative cases: i) one where 
these 
corrections 
are of IR origin and  of the ``standard'' $1/k^4$ type, and ii) one 
where one 
assumes 
unconventional (i.e. originating from new physics) $1/k^2$ contributions.

i) Let us first assume for the Euclidean quantity $D(Q^2)$:

\begin{equation}\Phi_D(k^2/Q^2)
\simeq  A\ {k^4\over Q^4} \label{eq:F-Dlow2}\end{equation}
at small $k^2$, corresponding to the standard dimension 4 ``gluon
condensate'', and put:

\begin{equation}\Phi_D(k^2/Q^2)
\equiv  A\ {k^4\over Q^4}+\Phi_D^{(3)}(k^2/Q^2) \label{eq:F-Dlow3}
\end{equation}
where 
\begin{equation}\Phi_D^{(3)}(k^2/Q^2)={\cal O}(k^6/ Q^6)\label{eq:k6}
\end{equation} 
at small $k^2$. On the
 other 
hand at large $k^2$ assume (as suggested by the ``standard'' model of 
eq.(\ref{eq:sol2}) below):

\begin{equation}\delta\bar{\alpha}_s(k^2)\simeq 
{c\over\log^2{k^2\over\Lambda^2}}
\ {\Lambda^4\over k^4}\label{eq:a-UV2}\end{equation}
Then we have:
\begin{equation}\delta D(Q^2)
=  \int_0^\infty{dk^2\over k^2}\ \delta\bar{\alpha}_s(k^2)\
A\ {k^4\over Q^4}+\int_0^\infty{dk^2\over k^2}\ 
\delta\bar{\alpha}_s(k^2)\
\Phi_D^{(3)}(k^2/Q^2)\label{eq:dD2}\end{equation}
where both integrals are UV convergent\footnote{The behavior eq.(\ref{eq:a-UV2})
is actually the ``hardest'' one allowing for an UV finite gluon condensate.}.
The first integral gives a ``gluon condensate'' contribution
of essentially IR origin (barring the high energy tail):
\begin{equation}\int_0^\infty{dk^2\over k^2}\ \delta\bar{\alpha}_s(k^2)\
A\ {k^4\over Q^4}= A\ K\ {\Lambda^4\over Q^4}\label{eq:dDIR1}
\end{equation}
where:
\begin{equation}K\equiv\int_0^\infty{dk^2\over k^2}\ 
\delta\bar{\alpha}_s(k^2)\
 {k^4\over \Lambda^4}\label{eq:K}\end{equation}
 On the other
hand, the second integral yields, at large $Q^2$, using 
eq.(\ref{eq:a-UV2}):
\begin{equation}\int_0^\infty{dk^2\over k^2}\ \delta\bar{\alpha}_s(k^2)\
\Phi_D^{(3)}(k^2/Q^2)\simeq c\ {\Lambda^4\over Q^4}
\int_0^{\infty}{dk^2\over k^2}\
{Q^4\over k^4}\ \Phi_D^{(3)}(k^2/Q^2)
\  {1\over\log^2{k^2\over\Lambda^2}}
\label{eq:dDUV1}\end{equation}
Noting that $\Phi_D(k^2/Q^2)$ must vanish at large $k^2$ (to insure UV 
convergence of the defining integral eq.(\ref{eq:D})), it follows that
${Q^4\over k^4}\ \Phi_D^{(3)}(k^2/Q^2)\rightarrow -A$ for 
$k^2\rightarrow\infty$, which implies the integral on the right hand side
 of 
eq.(\ref{eq:dDUV1}) is dominated by the UV region, and has the leading 
behavior at large $Q^2$:
\begin{equation}\int_0^{\infty}{dk^2\over k^2}\
{Q^4\over k^4}\ \Phi_D^{(3)}(k^2/Q^2)
\  {1\over\log^2{k^2\over\Lambda^2}}\simeq -\ {A\over \log{Q^2\over
 \Lambda^2}}
\label{eq:dDUV2}\end{equation}
Eq.(\ref{eq:dDUV2}) can be checked by splitting the integral at 
$k^2=Q^2$.
Note that the integral is IR convergent due to eq.(\ref{eq:k6}), and that
 one gets an ${\cal O}(1/\log Q^2)$ behavior, rather then an 
${\cal O}(1/\log^2 Q^2)$ one, reflecting the non-vanishing of 
${Q^4\over k^4}\ \Phi_D^{(3)}(k^2/Q^2)$ at large $k^2$. 
Hence:
\begin{equation}\int_0^\infty{dk^2\over k^2}\ \delta\bar{\alpha}_s(k^2)\
\Phi_D^{(3)}(k^2/Q^2)\simeq -A\ c\ {\Lambda^4\over Q^4}\ 
{1\over \log{Q^2\over \Lambda^2}}\label{eq:dDUV3}\end{equation}
I stress that this is an {\em ultraviolet} correction, 
insensitive to
 any
 IR cutoff $\mu_I^2$ one might introduce in the integral in 
eq.(\ref{eq:dDUV3}),
 since the low energy part of the integral below $\mu_I^2$ contributes a 
term 
much smaller then $1/Q^4$ at large $Q^2$. One therefore ends up with:
\begin{equation}\delta D(Q^2)\simeq  A\ K\ {\Lambda^4\over Q^4}\ 
\left(1-{c\over K}\
{1\over \log{Q^2\over \Lambda^2}}+...\right)\label{eq:dD3}\end{equation}
It is natural to interpret the logarithmically suppressed term
in eq.(\ref{eq:dD3}) as a contribution to the gluon condensate 
coefficient 
function. One then has to face an apparent paradox, since in QCD
 such a contribution is usually 
thought to arise from diagrams with {\em two}  gluon exchanges, with one 
soft and one hard gluon line, not considered in the present (dressed)
 {\em single} gluon exchange framework. I conclude that the gluon
 condensate
coefficient
function computed in the ``non-perturbative vacuum'' (where 
non-perturbative contributions to $\bar{\alpha}_s$ are taken into 
account)
 must differ from the 
``naive''  one computed with standard methods in the
``perturbative vacuum''. Alternatively,  one may adopt 
the convention to assign the new contribution to the 
{\em identity operator} coefficient function, where it would appear
as a  power suppressed  correction. Such a reshuffling will
preserve the property that at least the power series part of the
 coefficient functions is correctly given by the standard perturbation
theory approach. I stress that this unconventional contribution does not
 really mean new physics (at the 
difference
of the $1/Q^2$ terms to be discussed below), since it is 
generated by  a power suppressed
 term in $\bar{\alpha}_s$  of standard {\em infrared} origin. 
 The situation here looks 
similar
to the one  in the two dimensional ${\cal O}(N)$ non-linear 
$\sigma$-model,
which has been recently reanalysed in the $1/N$ expansion
in \cite{BBK}. It is thus  possible that also in the latter model
coefficients functions computed in the perturbative phase may 
differ\footnote 
{This possibility has not been checked in \cite{BBK}.}
from those computed in the true vacuum.

Eq.(\ref{eq:dD3}) poses another interesting theoretical
problem: one might expect the coefficient $c/K$ to be unambiguous, and 
all the ambiguity in $\delta D$ to reside in the overall normalization
 factor $K$ (eq.(\ref{eq:K})), the ``matrix element''. The latter is indeed 
ambiguous if 
$\bar{\alpha}_s^{PT}$, hence $\delta\bar{\alpha}_s$, have a (space-like)
Landau singularity. Within the present assumptions where $\bar{\alpha}_s^{PT}$
has no renormalons\footnote{It is difficult to see how the ambiguities in $K$
 and 
$c$ could possibly cancell if one makes the alternative
assumption that $\bar{\alpha}_s^{PT}$ has
renormalons.}
 and $c$ is consequently  unambiguous, $c/K$ appears
however to be ambiguous. One may think of several ways out of this possible
inconsistency:

a) One can introduce an IR cut-off $\mu_I$ in eq.(\ref{eq:dD2}) and write
eq.(\ref{eq:dD3}) as:
\begin{equation}\delta D(Q^2)\simeq  A\ K_{IR}(\mu_I)\ {\Lambda^4\over Q^4}+
A\ K_{UV}(\mu_I)\ {\Lambda^4\over Q^4}\ 
\left(1-{c\over K_{UV}(\mu_I)}\
{1\over \log{Q^2\over \Lambda^2}}+...\right)\label{eq:dD3a}\end{equation}
where:
\begin{eqnarray}K_{IR}(\mu_I)&\equiv&\int_0^{\mu_I^2}{dk^2\over k^2}\ 
\delta\bar{\alpha}_s(k^2)\
 {k^4\over \Lambda^4}\nonumber\\
K_{UV}(\mu_I)&\equiv&\int_{\mu_I^2}^\infty{dk^2\over k^2}\ 
\delta\bar{\alpha}_s(k^2)\
 {k^4\over \Lambda^4}\label{eq:KIR-UV}\end{eqnarray}
with $K_{IR}(\mu_I)+K_{UV}(\mu_I)=K$,
and only $K_{IR}(\mu_I)$ is ambiguous.

b) Alternatively, if one does not wish to introduce $\mu_I$, one can 
imagine a split:
\begin{equation}\delta\bar{\alpha}_s(k^2)=\delta\bar{\alpha}_s^{IR}(k^2)+
\delta\bar{\alpha}_s^{UV}(k^2)\label{eq:dsplit1}\end{equation}
where $\delta\bar{\alpha}_s^{IR}$ takes care of the Landau singularity
and is confined to the IR region, i.e. decreases very fast at large $k^2$, 
whereas $\delta\bar{\alpha}_s^{UV}$ contains no Landau singularity
 and behaves as $\delta\bar{\alpha}_s$ (eq.(\ref{eq:a-UV2})):
it is indeed possible to cancell the Landau singularity
with a $\delta\bar{\alpha}_s^{IR}$ contribution  
which
 decreases exponentially at large $k^2$ (see Appendix B).
Then one can 
write eq.(\ref{eq:dD3}) as:
\begin{equation}\delta D(Q^2)\simeq  A\ K_{IR}\ {\Lambda^4\over Q^4}+
A\ K_{UV}\ {\Lambda^4\over Q^4}\ 
\left(1-{c\over K_{UV}}\
{1\over \log{Q^2\over \Lambda^2}}+...\right)\label{eq:dD3b}\end{equation}
where:
\begin{eqnarray}K_{IR}&\equiv&\int_0^\infty{dk^2\over k^2}\ 
\delta\bar{\alpha}_s^{IR}(k^2)\
 {k^4\over \Lambda^4}\nonumber\\
K_{UV}&\equiv&\int_0^\infty{dk^2\over k^2}\ 
\delta\bar{\alpha}_s^{UV}(k^2)\
 {k^4\over \Lambda^4}\label{eq:KIR-UV1}\end{eqnarray}
with $K_{IR}+K_{UV}=K$,
and only $K_{IR}$ is ambiguous.

c) An attractive third alternative assumes
that $K$ is actually  unambiguous. This
is possible\footnote{A weaker, but more artificial, condition is to assume
 that 
$\bar{\alpha}_s^{PT}$ has only {\em complex} \cite{GGK} Landau singularities.}
 if $\bar{\alpha}_s^{PT}$ turns out to be 
 ``causal''
(section 3), which implies that $\bar{\alpha}_s^{PT}$
and $\delta\bar{\alpha}_s$ have no Landau singularities. Note however that
 in this case $\bar{\alpha}_s^{PT}=\bar{\alpha}_s^{APT}$, and 
consequently
$\int_0^\infty{dk^2\over k^2}
\ \bar{\alpha}_s^{PT}(k^2)\
\Phi_D(k^2/Q^2)=D_{APT}(Q^2)$ differs from the corresponding Borel sum
$D_{PT}(Q^2)$ by power terms $\delta D_{APT}(Q^2)$. In leading order we 
have, since the coupling is causal (see Appendix B):
\begin{equation}\delta D_{APT}(Q^2)\simeq  A\ K_{APT}\ {\Lambda^4\over Q^4}
+...\label{eq:dD-APT}\end{equation}
where $K_{APT}$ is ambiguous (it cancells the renormalon ambiguity still
present \cite{G4,DU} in $D_{PT}(Q^2)$). Therefore we end up with:
\[D(Q^2)=D_{PT}(Q^2)+\delta D(Q^2)\]
where:
\begin{eqnarray}\delta D(Q^2)&\equiv&\delta D_{APT}(Q^2)+\delta D_{NP}(Q^2)
\nonumber\\
&\simeq& A\ K_{APT}\ {\Lambda^4\over Q^4}+
A\ K_{NP}\ {\Lambda^4\over Q^4}\ 
\left(1-{c\over K_{NP}}\
{1\over \log{Q^2\over \Lambda^2}}+...\right)
\label{eq:dD-tot}\end{eqnarray}
where $\delta D$ and $K$ in eq.(\ref{eq:dD3})
 have been renamed $\delta D_{NP}$ and $K_{NP}$.
Note that cases a) and b) are just arbitrary definition-dependent rewritings of
eq.(\ref{eq:dD3}), without new physical content. However, all three cases
suggest the general ansatz for the ${\cal O}(1/Q^4)$ contribution to have
the following 
``two-component'' form, once ``standard'' corrections due to double
gluon exchange are taken into account:
\begin{eqnarray}\delta D(Q^2)&\simeq&
A\ K_{PT}\ {\Lambda^4\over Q^4}\left(1+d_{PT}\
{1\over \log{Q^2\over \Lambda^2}}+...\right)\nonumber\\
&+& A\ K_{NP}\ {\Lambda^4\over Q^4}\ 
\left(1+(d_{PT}+d_{NP})\
{1\over \log{Q^2\over \Lambda^2}}+...\right)\label{eq:dD-tot1}\end{eqnarray}
where the first  contribution contains the ambiguity through the constant
$K_{PT}$ and has the standard perturbative coefficient $d_{PT}$, while
the second contribution arises in the non-perturbative vacuum, with
$d_{NP}=-c/K_{NP}$. Since $K_{PT}$
cancells the renormalon ambiguity, which depends on information contained in
 perturbation theory,
 the first contribution can be thought of
 being of a ``perturbative'' nature, at the difference of the second one, which 
 depends on a more genuinely  non-perturbative information. The different
logarithmic corrections in eq.(\ref{eq:dD-tot1}) should allow to separate 
unambiguously the two contributions.

ii) Let us next consider the case where there is a  leading 
 ${\cal O}(1/Q^2)$ power correction   of 
UV origin. 
 For the Euclidean quantity $D(Q^2)$,
 this means \cite{G2} that 
$|\Phi_D(k^2/Q^2)|\ll|{k^2\over Q^2}|$
at small $k^2$, while $\delta\bar{\alpha}_s(k^2)$ is ${\cal O}(1/k^2)$
at large $k^2$, so that the leading IR power correction is parametrically 
suppressed compared to the UV one. For instance, if one assumes:

\begin{equation}\delta\bar{\alpha}_s(k^2)\simeq b_1\ {\Lambda^2\over
k^2}\label{eq:aUV1}\end{equation}
and substitute into $\delta D_{UV}$ (eq.(\ref{eq:Duv-split})), one
gets at large $Q^2$:
\begin{equation}\delta D_{UV}(Q^2)
\simeq  A_D\ b_1\ {\Lambda^2\over Q^2} \label{eq:dD1}\end{equation}
with:
\begin{equation}A_D\equiv\int_0^\infty{dk^2\over k^2}\ {Q^2\over k^2}\
\Phi_D(k^2/Q^2)\label{eq: A_D}\end{equation}
where the integral is IR convergent from the stated assumptions. Note the
 same assumption on $\Phi_D$ implies,  
expanding at small $\mu^2$ under the integral in eq.(\ref{eq:FDdisp}):
\begin{equation}\dot{{\cal F}}_D(\mu^2/Q^2)
\simeq  A_D\ {\mu^2\over Q^2} \label{eq:F-Dlow}\end{equation}
i.e. an {\em analytic\/} small $\mu^2$ behavior.

The remarks in \cite{G2} can be 
generalized to Minkowskian quantities $R(Q^2)$. Let us similarly 
assume for small $\mu^2$:
\begin{equation}\dot{{\cal F}}_R(\mu^2/Q^2)
\simeq  A_R\ {\mu^2\over Q^2} \label{eq:F-Rlow}\end{equation}
while, at large $k^2$, allowing for a logarithmic correction for the sake
 of generality:
\begin{equation}\delta\bar{\alpha}_s(k^2)\simeq 
\left(b_1+{c_1\over\log{k^2\over\Lambda^2}}\right)
{\Lambda^2\over
k^2}\label{eq:aUV}
\end{equation}
where $b_1$ and $c_1$ are non-perturbative parameters (this is
 only an illustrative example, since no
theory presently exists for these ``unorthodox'' power corrections).

A derivation of the analogue of eq.(\ref{eq:dD1})
 based on the results of Appendix C (e.g. eq.(\ref{eq:dR}),
or eq.(\ref{eq:dR2})) is possible, but cumbersome, since one has to
introduce the split eq.(\ref{eq:asplit1}), and also
write down  dispersion relations which usually involve subtraction
constants, whereas only the assumption eq.(\ref{eq:F-Rlow}) is really
needed. In particular, the constant $A_R$ is equal to the subtraction
constant $a_0$, and is independent of $\Phi_R$,
 if  eq.(\ref{eq:Fdisp2}) is assumed; then
the leading UV power correction is given by the first, ``subtraction''
term on the right hand side of eq.(\ref{eq:dR2}), which yields a 
result similar to eq.(\ref{eq:dD1}). Alternatively, one could 
use \cite{G1}
the expression for power corrections in term of the Minkowkian coupling
$\bar{\alpha}_{eff}$ (see Appendix B), but this approach is  
unconvenient too, since it again relies  on the 
split eq.(\ref{eq:asplit1}). 

All these problems are however
 circumvented if one 
deals first with the  associated Euclidean quantity
$D(Q^2)$, and deduce the corresponding power corrections for $R(Q^2)$
using analyticity (eq.(\ref{eq:dR-dDreg})). Indeed eq.(\ref{eq:ker}) and
 eq.(\ref{eq:F-Rlow}) imply the {\em analytic} small $k^2$ behavior of 
the Euclidean
kernel $\Phi_D$:
\begin{equation}\Phi_D(k^2/Q^2)
\simeq  A_R\ {k^2\over Q^2} \label{eq:Fi-Dlow}\end{equation}
(the corresponding $\dot{{\cal F}}_D$ behaves as $A_R\ {\mu^2\over Q^2}
\log{Q^2\over\mu^2}$ at small $\mu^2$, and signals the contribution
of a d=2 operator
 in the OPE of the considered Euclidean correlation 
function: this is why eq.(\ref{eq:F-Rlow}) does not hold for e.g.
$R_{e^+e^-}$).
One can then show (see Appendix A), at large $Q^2$:
\begin{equation}\delta D_{UV}(Q^2)\simeq{\Lambda^2\over Q^2}
\left[A_R \left(b_1\log{Q^2\over\Lambda^2}+c_1\log\log{Q^2\over\Lambda^2}
\right)+ const + {\cal O}\left({1\over\log{Q^2\over\Lambda^2}}\right)
\right]
\label{eq:D-uv}\end{equation}
where the  constant and ${\cal O}(1/\log Q^2)$ terms cannot be expressed
 only
in term of the  small $k^2$ behavior of $\Phi_D$, and the first two 
terms on the right-hand side represent also the leading contributions to
$\delta D$, and are of {\em ultraviolet\/} origin ($\delta D$ 
contains in addition an ${\cal O}(1/Q^2)$ piece of IR origin,
 coming from the first
term in eq.(\ref{eq:dDsplit})) .
 Applying the operator 
${1\over 2\pi i} \oint_{|Q'^2|=Q^2}{dQ'^2\over Q'^2}$ to the right-hand 
side of 
eq.(\ref{eq:D-uv}) then yields, as a consequence of 
eq.(\ref{eq:dR-dDreg}),
 the leading contribution to 
$\delta R_{UV}$ (and  $\delta R$)
at large $Q^2$  :
\begin{equation}\delta R_{UV}(Q^2)\simeq
{\Lambda^2\over Q^2}\left[A_R\left(b_1+{c_1\over\log{Q^2\over\Lambda^2}}
\right)+{\cal O}\left({1\over\log^2{Q^2\over\Lambda^2}}\right)\right]
\label{eq:R-uv}\end{equation}
where the   ${\cal O}(1/\log^2 Q^2)$ term cannot be expressed
 only
in term of the  small $\mu^2$ behavior of $\dot{{\cal F}}_R$.
As a check, eq.(\ref{eq:D-uv}) can also be obtained directly from 
eq.(\ref{eq:R-uv})
and the dispersion relation eq.(\ref{eq:D-R}) using similar methods as 
in Appendix A. Eq.(\ref{eq:R-uv}), which is the analogue of 
eq.(\ref{eq:dD1}),  is naturally identified as an 
{\em ultraviolet\/} contribution; note in this respect that the leading 
IR power contribution originating
from the term $D_{IR}(Q^2)$ in eq.(\ref{eq:D-split}) is  a pure power, 
if
$\Phi_D$ is analytic at small $k^2$,  hence
gives no contribution to $R$ after applying the operator
${1\over 2\pi i} \oint_{|Q'^2|=Q^2}{dQ'^2\over Q'^2}$.

One can then relate simply the UV power corrections in  different
channels, if the non-perturbative parameters $b_1$ and $c_1$,
which have to be fitted, are assumed
to be {\em universal} (like the ``physical'' coupling $\bar{\alpha}_s$
itself), since the channel-dependant parameter $A_R$ is calculable from
Feynman diagrams. These remarks lead to a simple phenomenology \cite{G2}
 of
$1/Q^2$ terms (if they turn out large enough to be detected). The channel
dependence may be sizable.
 For instance, writing a generic Euclidean correlation function computed
 at one loop with a  gluon of mass $\mu^2$ as\footnote{The normalization
of ${\cal F}_D$ here is different from the one in eq.(\ref{eq:D1}).}:
\begin{equation}D(Q^2)=1+{\alpha_s\over \pi}\left[{\cal F}_D(0)-A_D\ 
{\mu^2\over Q^2}+...\right] \label{eq:D-g}\end{equation}
(where the parton model contribution is normalized to unity), 
one has \cite{BBB}:
\begin{equation}A_{vector}={8\over 3}(4-3\zeta (3)) \simeq 1.07
\label{eq:A-V}\end{equation}
for the vector correlation function, whereas \cite{B3}:
\begin{equation}A_{pseudoscalar}=4\label{eq:A-P}\end{equation}
for the pseudoscalar correlation 
function\footnote{A closely related result is quoted in \cite{Z1,CNZ}.}
 (after taking two derivatives
\cite{P}
to get rid of an overall UV divergence and extracting two powers of quark
 masses). It
is thus possible that eventual $1/Q^2$ terms (whatever their physical
origin) may have more important effects in the pseudoscalar channel 
\cite{CNZ} then
 in the vector one (where they seem  to be negligible \cite{Nar}).
 Unfortunately, a
phenomenological determination of such terms may be difficult given the 
large {\em perturbative} corrections found in the pseudoscalar channel.

\noindent\underline{Currents with anomalous dimensions}: in the 
pseudoscalar
case (and in many other), the corresponding current has anomalous 
dimension.
 Here  I suggest an ansatz  to deal with the latter. It is useful to 
factor out explicitly the anomalous dimension dependence using the 
renormalization group, and write the correlation function as:
\begin{equation} D(Q^2)={D_{inv}(Q^2)\over Z(\mu^2)}\label{eq:D-P}
\end{equation}
where $D_{inv}(Q^2)$ is 
renormalization group invariant, and $Z(\mu^2)$ is the integrated 
anomalous
dimension factor, related to the anomalous dimension function 
$\gamma(\alpha_s)$ by:
\begin{equation}{d\log Z\over d\log\mu^2}= 
\gamma\left(\alpha_s(\mu^2)\right)
\label{eq:Z}\end{equation}
In perturbation theory, we have:
\begin{equation}D_{inv}^{PT}(Q^2)=
(\alpha_s)^{\gamma_0/\beta_0}\left(1+{\cal O}
(\alpha_s)\right)\label{eq:D-invpt}\end{equation}
where $\gamma_0$ is the one loop anomalous dimension, 
and similarly for $Z(\mu^2)$. It is convenient to introduce the 
``effective charges'' \cite{Grunberg}:
\begin{equation}D_{inv}(Q^2)\equiv \left[A(Q^2)\right]^{\gamma_0/\beta_0}
\label{eq:Abar-eff}\end{equation}
and similarly
\begin{equation}Z(\mu^2)\equiv \left[a(\mu^2)\right]^{\gamma_0/\beta_0}
\label{eq:A-eff}\end{equation}
where both ${A}(Q^2)$ and $a(\mu^2)$ are ${\cal O}(\alpha_s)$ quantities.
 Then:
\begin{equation}D(Q^2)=
\left[{A(Q^2)\over a(\mu^2)}\right]^{\gamma_0/\beta_0}
\label{eq:D-inv}\end{equation}
Let us now assume that an ${\cal O}(1/Q^2)$ 
correction appears in the ``non-perturbative''
 RG invariant correlation function in the form:
\begin{equation}D_{inv}(Q^2)=D_{inv}^{PT}(Q^2)\left(1 +{\gamma_0\over
 \beta_0}
\ {C\over Q^2}+...\right)\label{eq:D-invnp}\end{equation}
which  implies:
\begin{equation}A(Q^2)=A_{PT}(Q^2)\left(1 +{C\over Q^2}+...\right)
\label{eq:A-invnp}\end{equation}
whereas no such corrections are expected in 
$Z(\mu^2)$ and $a(\mu^2)$, which are entirely perturbative quantities. 
Then we have at large $Q^2$:
\begin{equation}\log\left[{A(Q^2)\over a(\mu^2)}\right]=
\log\left[{A_{PT}(Q^2)\over a(\mu^2)}\right]+{C\over Q^2}+...
\label{eq:logA-np}\end{equation}
Note that $\log\left[{A_{PT}(Q^2)\over a(\mu^2)}\right]$ is an
${\cal O}(\alpha_s)$ quantity 
\footnote{It is interesting to note that 
$\log\left[A_{PT}(Q^2)\right]$ corresponds formally to the image in
 coupling
constant space of the ``bare Borel transform'' of \cite{BBB}: the
logarithmic
dependence on the coupling simply reflects the singular behavior of the
``bare Borel transform'' at the origin. Similarly, 
$\log\left[{A_{PT}(Q^2)\over a(\mu^2)}\right]$ is essentially
the image of the corresponding
renormalized Borel transform (see Appendix A in \cite{BBB}).}.
 In order to make contact between
this quantity and the single gluon
exchange integral, I now appeal to the large $N_f$ limit, which displays
 the 
corresponding single renormalon chain type of diagrams. In this
limit eq.(\ref{eq:D-inv}) yields, since $\gamma_0/\beta_0$ is
 ${\cal O}(1/N_f)$:
\begin{equation}D(Q^2)=1+{\gamma_0\over\beta_0} 
\log\left[{A(Q^2)\over a(\mu^2)}\right]_{\vert{N_f=\infty}}+
{\cal O}(1/N_f^2)
\label{eq:D-Nf}\end{equation}
which shows that 
${\gamma_0\over\beta_0} \log\left[{A_{PT}(Q^2)\over a(\mu^2)}
\right]_{\vert{N_f=\infty}}$
should be identified to the sum of the single renormalon chain diagrams. 
Consequently, the power correction eq.(\ref{eq:dD1}) obtained from these 
diagrams should be identified to ${\gamma_0\over\beta_0} {C\over Q^2}$,
 i.e. we
have:
\begin{equation}{\gamma_0\over\beta_0}\ C=A_D\ b_1\label{eq:C}
\end{equation}

\section{Ans\"atze for the non-perturbative contributions to the
universal QCD running
coupling}
\subsection{QED inspired models} Although they are usually completely 
neglected, 
power suppressed 
corrections to a ``physical'' coupling such as 
$\bar{\alpha}_s$, which is supposed to be defined also at the 
non-perturbative level (and is assumed in \cite{DMW} to be {\em the} 
universal  coupling of QCD) are a natural expectation, and could 
eventually
 be derived from the OPE
itself as the following QED analogy shows. In QED, the coupling
$\bar{\alpha}_s(k^2)$ should be identified, in the present dressed
single gluon exchange context, to the Gell-Mann-Low effective charge
$\bar{\alpha}$, related to the photon vacuum polarisation $\Pi (k^2)$
by:
\begin{equation}\bar{\alpha}(k^2)= {\alpha\over 1+ \alpha\
\Pi(k^2/\mu^2,\alpha)}\label{eq:a-qed}\end{equation}
One  expects  $\Pi(k^2)$, hence $\bar{\alpha}(k^2)$,  to receive power
contributions from the OPE. Of course, this cannot happen in QED
itself, which is an IR trivial theory, but might occur in the ``large
$\beta_0$'', $N_f=-\infty$ limit of QCD . Instead of $\Pi(k^2)$, it is
convenient to introduce the related (properly normalized)
renormalization group invariant ``Adler function'' (with the Born term
removed):
\begin{equation}{\cal A}(k^2)= {1\over \beta_0}\left({d\Pi\over d\log
k^2}-{d\Pi\over d\log k^2}|\alpha=0\right)\label{eq:adler}\end{equation}
${\cal A}_s$, its assumed analogue in QCD,
 contributes the higher order terms in the renormalization group
equation:
\begin{equation}{d\bar{\alpha}_s\over d\ln k^2} = -\beta_0
(\bar{\alpha}_s)^2\left(1+  {\cal A}_s\right)\label{eq:RG}\end{equation}

Consider now the  $N_f=-\infty$ limit in QCD. Then ${\cal A}_s(k^2)$ is 
expected
to be purely non-perturbative, since in this limit the perturbative
part of $\bar{\alpha}_s$ is just the one-loop coupling
$\bar{\alpha}_s^{PT}(k^2)=1/\beta_0\ln(k^2/\Lambda^2)$.
Indeed, OPE-renormalons type arguments suggest the general structure
\cite{G-qed} at large $k^2$:
\begin{equation}{\cal A}_s(k^2)=\sum_{p=1}^\infty \left(a_p \log{k^2\over
\Lambda^2}+ b_p\right) \left(\ {\Lambda^2\over
k^2}\right)^p\label{eq:R-np}\end{equation}
where the log enhanced power corrections reflect the presence of double
IR renormalons poles \cite{B4}. Eq.(\ref{eq:RG}) with ${\cal A}_s$ as in 
eq.(\ref{eq:R-np}) can  be
easily integrated
 to give:
\begin{equation}{1\over\beta_0\bar{\alpha}_s}=\log{k^2\over\Lambda^2}-
\sum_{p=1}^\infty {1\over p}\left(\ {\Lambda^2\over k^2}\right)^p 
\left(a_p \log{k^2\over\Lambda^2}+ b_p+{a_p\over p}\right)
\label{eq:sol}\end{equation}
which yields the expansion:

\begin{equation}\bar{\alpha}_s(k^2)=\bar{\alpha}_s^{PT}(k^2)+
\sum_{p=1}^\infty \left[c_{p,0}\ \bar{\alpha}_s^{PT}(k^2)+....+
c_{p,p}\left(\bar{\alpha}_s^{PT}(k^2)\right)^{p+1}\right] 
\left(\ {\Lambda^2\over
k^2}\right)^p\label{eq:sol1}\end{equation}
where the first terms are given by:

\[\left\{\begin{array}{lll}c_{1,0}&=&a_1\\
c_{1,1}&=&\beta_0\ (a_1+b_1)\end{array}\right.\]

and:
\[\left\{\begin{array}{lll}c_{2,0}&=&{1\over 2}a_2+a_1^2\\
c_{2,1}&=&\beta_0\ ({1\over 4}a_2+{1\over 2}b_2+2a_1b_1+2a_1^2)\\
c_{2,2}&=&(\beta_0)^2\ (a_1+b_1)^2\end{array}\right.\]

In QCD, one actually expects : $a_1=b_1=0$ (reflecting the absence of
$d=2$ gauge invariant  operator), and $a_2=0$ (reflecting the
absence of anomalous dimension in the gluon condensate), which
gives:
\begin{equation}\bar{\alpha}_s(k^2)=\bar{\alpha}_s^{PT}(k^2)
+{1\over 2}\beta_0\ b_2\ \left(\bar{\alpha}_s^{PT}(k^2)
\right)^2\ {\Lambda^4\over k^4} +...\label{eq:sol2}\end{equation}
i.e. in this semi-standard framework there is no
(as expected) $1/k^2$ correction.
It is interesting to note that keeping only the $p=2$ (gluon condensate)
contribution in eq.(\ref{eq:R-np}) with $a_2=0$ yields:

\begin{equation}\bar{\alpha}_s(k^2)=
{1\over\beta_0\left(\ln{k^2\over\Lambda^2}-
{b_2\over 2}{\Lambda^4\over k^4}\right)}\label{eq:grib}\end{equation}
This model\footnote{It is amusing to note that the popular
Richardson coupling:
$\bar{\alpha}_s(k^2)=1/\beta_0 \log\left(c^2+k^2/\Lambda^2\right)$
 is also very simply described by  equation (\ref{eq:RG}).
It corresponds to:
${\cal A}_s(k^2)=-1/\left (1+k^2/c^2\Lambda^2\right)$
i.e. to  ${\cal A}_s$ given by a simple pole in the
time-like region, with prescribed residue.}
 coincides with a  previously suggested ansatz \cite{Grib}  based
 on different arguments, which lead to the suggestion that the running 
coupling should satisfy the second order differential equation:
\begin{equation}{1\over 2} \left(1\over \bar{\alpha}_s\right)''+
\left(1\over \bar{\alpha}_s\right)'=\beta_0\label{eq:grib1}\end{equation}
(where $'\equiv d/d\log k^2$)
whose solution turns out to be given by eq.(\ref{eq:grib}). A possibly welcome 
feature \cite{Dok} of the  coupling eq.(\ref{eq:grib}) is that it vanishes at
$k^2=0$.
However, it  probably also has Landau
 singularities on the first 
sheet of the complex $k^2$ plane, and does not satisfy the dispersion 
relation eq.(\ref{eq:disp2}).

It is tempting to speculate that the series in eq.(\ref{eq:R-np}) has a 
finite convergence radius. A general ansatz for $\bar{\alpha}_s$ would
then be given by the solution $\bar{\alpha}_s^{UV}$ of eq.(\ref{eq:RG})
 (i.e. the coupling
as reconstructed from its short distance expansion), augmented by a term
$\delta\bar{\alpha}_s^{IR}$ whose support is entirely in the infrared 
region 
(i.e. has an exponentially supressed UV tail),
both pieces being assumed to satisfy the dispersion relation
eq.(\ref{eq:disp2}):
\begin{equation}\bar{\alpha}_s(k^2)=\bar{\alpha}_s^{UV}(k^2)+
\delta\bar{\alpha}_s^{IR}(k^2)\label{eq:asplit-uvir}\end{equation}
 If it were possible to know $\bar{\alpha}_s^{UV}$
analytically, then the introduction of an IR cut-off $\mu_I$ in
 eq.(\ref{eq:D-split2}) would not be anymore necessary,\footnote{If it 
further happens that
$\bar{\alpha}_s^{UV}$ has its support in the ultraviolet region and 
is exponentially suppressed in the {\em infrared}, for instance
$\bar{\alpha}_s^{UV}(k^2)=\exp\left[-(\Lambda^2/k^2)^N\right]
1/\beta_0 \log\left(c^2+k^2/\Lambda^2\right)$ (this example satisfies
eq.(\ref{eq:RG}) and (\ref{eq:R-np})),
 then the ansatz eq.(\ref{eq:asplit-uvir})
would be a natural and unique smooth-out substitute of the 
introduction of a sharp IR cut-off (eq.(\ref{eq:D-split2})), which 
amounts to write:
$\bar{\alpha}_s=\bar{\alpha}_s \theta(k^2-\mu_I^2)+
\bar{\alpha}_s \theta(\mu_I^2-k^2)$
where the first and second term play the role respectively of 
$\bar{\alpha}_s^{UV}$ and $\delta\bar{\alpha}_s^{IR}$. Taking e.g.
$\delta\bar{\alpha}_s^{IR}(k^2)=C\exp\left(-k^2/\Lambda^2\right)$
(this choice 
 is not quite satisfactory, since it violates asymptotic freedom for
 $Re(k^2)<0)$), 
eq.(\ref{eq:asplit-uvir})
reproduces (for $C=-1$ and $N=1$) an example given in \cite{P} 
(with a 
different motivation).}
 and one could 
parametrize the remaining IR contributions  with (UV convergent) moments
 of the $\delta\bar{\alpha}_s^{IR}$ piece, along the lines of \cite{DMW}.

\subsection{Models based on the ``analytic perturbation theory''
coupling}

The ``analytic perturbation theory'' (APT) coupling \cite{SS,G1,GGK,BBB}:
\begin{equation}\bar{\alpha}_s^{APT}(k^2)
\equiv k^2\int_0^\infty{d\mu^2
\over(\mu^2+k^2)^2}\
\bar{\alpha}_{eff}^{PT}(\mu^2)\label{eq:APT}\end{equation}
 is
used in Appendices B and C for entirely technical reasons, to deal
with power corrections in the ``Minkowskian representation''. It is 
nevertheless tempting to speculate about its eventual physical 
relevance
(some remarkable infrared properties have been pointed out 
in \cite{SS,GGK}).
 Since it is always possible to 
perform the split eq.(\ref{eq:asplit1}):
\[\bar{\alpha}_s(k^2)=\bar{\alpha}_s^{APT}(k^2)+
\delta\bar{\alpha}_s^{ANP}(k^2)\]
 which is just a definition,
 physical content arises only if 
additional assumptions are made concerning $\delta\bar{\alpha}_s^{ANP}$.
Two simple possibilities\footnote{For another suggestion 
see \cite{Webber}.} come to mind:

i) One can assume \cite{G1} that $\delta\bar{\alpha}_s^{ANP}(k^2)\equiv
\delta\bar{\alpha}_s^{IR}(k^2)$ is 
an essentially infrared contribution, highly
suppressed in the {\em ultraviolet} region: $\bar{\alpha}_s^{APT}$
thus plays the role\footnote{However the corresponding 
${\cal A}_s(k^2)$ 
does not have the 
structure of eq.(\ref{eq:R-np}).}
 of $\bar{\alpha}_s^{UV}$ in 
eq.(\ref{eq:asplit-uvir}). 
This assumption leads, {\em unless $\bar{\alpha}_s^{PT}$ is causal}, to the 
prediction of unconventional, OPE unrelated $1/Q^2$ power terms, which 
(when calculated with a one-loop model)  seem to predict \cite{G1} a too 
large
value of $\alpha_s$ in $\tau$-decay. One also obtains the wrong 
sign \cite{AZ} for the detected \cite{Marchesini} $1/Q^2$ term to the 
lattice 
gluon
condensate. These problems are avoided, as mentionned above, if one assumes
that $\bar{\alpha}_s^{PT}$ is causal, i.e. that 
$\bar{\alpha}_s^{PT}=\bar{\alpha}_s^{APT}$, which is a natural 
possibility
 in the framework of \cite{DMW} (see also the discussion in section 5).

ii) Alternatively, one can make the opposite assumption that 
$\delta\bar{\alpha}_s^{ANP}(k^2)
\equiv\delta\bar{\alpha}_s^{UV}(k^2)$ is 
an essentially ultraviolet contribution, highly
suppressed in the {\em infrared} region (e.g. 
$\delta\bar{\alpha}_s^{UV}(k^2)=C\ \exp\left[-(\Lambda^2/k^2)^N\right]
{\Lambda^2\over k^2}$). This assumption implies that the {\em infrared}
part of the total  $\bar{\alpha}_s$ coupling is essentially given by that
of
$\bar{\alpha}_s^{APT}$, while making no commitment upon the 
magnitude and sign of eventual $1/Q^2$ terms. Furthermore, 
it turns out that the low energy behavior of the APT
coupling is rather insensitive \cite{SS,GGK} to higher loops in the
perturbative beta function, and close to
 the one-loop APT model (eq.(\ref{eq:apt-1loop})),
 for a large class
of renormalization schemes (the important exception \cite{GGK} to the previous 
statement is again the case where
 $\bar{\alpha}_s^{PT}$ is causal). Thus, unless  $\bar{\alpha}_s^{PT}$ is 
causal,
 $R_{IR}(Q^2)$ (eq.(\ref{eq:Rir})) 
will
be approximated by:
\begin{equation}R_{IR}(Q^2)\simeq \int_0^{\mu_I^2}{dk^2\over k^2}
\ \bar{\alpha}_s^{APT}(k^2)\ \Phi_R(k^2/Q^2)
\simeq \int_0^{\mu_I^2}{dk^2\over k^2}
\ \bar{\alpha}_s^{APT}(k^2)_{\vert one-loop}\ \Phi_R(k^2/Q^2)
\label{eq:Rir-APT}\end{equation}
i.e. an essentially {\em parameter free} (apart from $\mu_I$) prediction~!
There may be some 
preliminary
phenomenological evidence \cite{SS} in favor of such an assumption (see
however \cite{Alek} for a possible theoretical inconsistency).

\section{Scheme dependence issues}
In standard applications, one neglects UV power corrections, and uses
 eq.(\ref{eq:R-split}):
\begin{equation}R(Q^2)\simeq R_{PT}(Q^2)+\int_0^{\mu_I^2}{dk^2\over k^2}
\ \bar{\alpha}_s(k^2)\ \Phi_R(k^2/Q^2)-
\int_0^{\mu_I^2}{dk^2\over k^2}
\ \bar{\alpha}_s^{PT}(k^2)\ \Phi_R(k^2/Q^2)
\label{eq:R-standard}\end{equation} 
where  $R_{PT}(Q^2)$ is now meant to be the full available
 (usually next to 
leading order) perturbative QCD expression for $R$, and is not restricted
to be given by the single gluon exchange expression eq.(\ref{eq:Rb})
(contrary to the two power corrections integrals in 
eq.(\ref{eq:R-standard})). 
This procedure however  suffers from the usual scheme dependence
ambiguity concerning the choice of renormalization scheme (RS) and scale 
parameter in the trunkated expression for $R_{PT}(Q^2)$. This problem
becomes  severe if next to leading order corrections are
large with the usual choice of RS (this happens in particular in the 
case of thrust \cite{DW}, where these corrections in the $\overline{MS}$
scheme with the standard choice of scale are close to 40\% at the Z 
mass). A resummation of such large corrections appears necessary. One
reasonable procedure for doing so, in absence of any other information,
 is the ``effective charge'' scheme \cite{Grunberg}. It has recently 
been applied \cite{CGM} to the case of thrust, where it yields results 
similar to those obtained with a ``low $\mu$'' choice \cite{Beneke} of 
renormalization point, close to the one ($\mu=0.08 Q$) which sets to zero 
the next to 
leading order corrections. The effect of these alternative  procedures 
is to 
drastically reduce the size of the $1/Q$ power term needed to fit the 
data. It is clearly an important issue to determine the correct RS.

However, one should  remark that if {\em one sticks to the single
gluon exchange picture even for $R_{PT}(Q^2)$}, there is no real scheme
dependence issue anymore, since the physical universal running coupling
$\bar{\alpha}_s^{PT}(k^2)$ which dresses the gluon propagator
 used under the single gluon exchange integral
 has already been uniquely identified (up to next to leading order in the
$\overline{MS}$ scheme). This is clear\footnote{An  analoguous
 observation
 has been made before \cite{Lu} as the essential RS invariance of the QED
``dressed skeleton expansion''; it is also closely related to the 
intuition behind the BLM scheme \cite{BLM}.}
 for an Euclidean quantity, where
 one can write (eq.(\ref{eq:D-split})), neglecting UV power corrections:
\begin{equation}D(Q^2)\simeq \int_{0}^{\mu_I^2}{dk^2
\over k^2}\
\bar{\alpha}_s(k^2)\ \Phi_D(k^2/Q^2)+
\int_{\mu_I^2}^\infty{dk^2\over k^2}
\ \bar{\alpha}_s^{PT}(k^2)\
\Phi_D(k^2/Q^2)\label{eq:D-single}\end{equation}
and $\bar{\alpha}_s^{PT}(k^2)$ can be unambiguously determined in an RS 
invariant manner (in term of e.g. $\Lambda_{\overline{MS}}$) for all
 scales above the IR cut-off $\mu_I$ by 
integrating its own renormalization group equation (known presently only
 up to the first two (universal) loops). Note that the IR cut-off
$\mu_I$ alleviates the problem of integrating over the Landau 
singularity in the second integral 
(the ``regularized perturbation theory'' piece).

For Minkowskian quantities, where an integral representation in term of
 $\bar{\alpha}_s(k^2)$ is not available, one has to use the 
(RS invariant) Borel 
transform formalism of section 3, i.e. write (eq.(\ref{eq:R-split})):
\begin{eqnarray}R(Q^2)&\simeq& \int_{0}^{\mu_I^2}{dk^2
\over k^2}\
\bar{\alpha}_s(k^2)\ \Phi_R(k^2/Q^2)\nonumber\\
&+&
\int_0^\infty 
dz\ \tilde{\alpha}_{eff}(z)\ 
\left[\int_0^{\infty}{d\mu^2\over \mu^2}\
\dot{{\cal F}}_{R,UV}(\mu^2,Q^2)\ \exp
\left(-z\beta_0\ln{\mu^2\over
\Lambda^2}\right)\right]\label{eq:R-single}\end{eqnarray}
where the second integral is the perturbative part of the Beneke-Braun
like ``gluon mass'' integral eq.(\ref{eq:Raptuv}) (but with the 
``IR regularized'' characteristic function replacing the full
characteristic function).
It would be interesting to establish the effect \cite{G.Grunberg} of  
using eq.(\ref{eq:R-single}) on the size of the $1/Q$ term in the 
thrust case. The crucial question underlying the reliability of this 
method is whether the alleged QCD
``dressed skeleton expansion'', trunkated at its first term (the single
gluon exchange level) is a good approximation.

\section{Summary and Conclusions}
Power corrections to generic QCD Minkowskian observables have been 
discussed in the 
``dressed single gluon exchange approximation'', assuming the existence 
of a 
(universal) QCD coupling $\bar{\alpha}_s$ defined at the non-perturbative
 level, and
regular in the infrared region. Following \cite{DW}, one introduces 
an IR
cut-off $\mu_I$, and distinguish
between  IR and UV power contributions.
This procedure appears hardly avoidable in practice:
 the alternative approach based on the split
eq.(\ref{eq:asplit-uvir}) $\bar{\alpha}_s=\bar{\alpha}_s^{UV}+
\delta\bar{\alpha}_s^{IR}$ (where $\delta\bar{\alpha}_s^{IR}$ is 
restricted to the infrared region)
 requires 
a reconstruction of the ``correct'' unique short distance coupling 
$\bar{\alpha}_s^{UV}$ from the short distance expansion of 
$\bar{\alpha}_s$, which is very 
hard to achieve (assuming it can be done at all), since one may have to sum
an infinite set of power suppressed corrections:
$\bar{\alpha}_s^{UV}$ cannot be given only by the perturbative part 
$\bar{\alpha}_s^{PT}$  which has a Landau singularity, unless 
$\bar{\alpha}_s^{PT}$ is causal. 
The adopted procedure 
allows a 
parametrization of IR power corrections in terms of low energy moments
of the full coupling $\bar{\alpha}_s$, which form one set of (universal)
non-perturbative parameters. On the other hand, UV power corrections 
rely on the split
of eq.(\ref{eq:asplit}) between perturbative and non-perturbative 
contributions
to $\bar{\alpha}_s$: the other set of (universal)
non-perturbative parameters are those which occur 
(assuming 
 a simple form) in the high
energy expansion of the non-perturbative part $\delta\bar{\alpha}_s$, 
i.e. power corrections to the coupling itself. I argued that the latter
are a natural expectation for a coupling such as $\bar{\alpha}_s$ assumed
to be defined at the non-perturbative level, and discussed some simple
 models for them which do not necessarily imply new physics. However, I argued 
that in general ``physical'' models for the universal coupling are not expected
to have too highly suppressed corrections to asymptotic freedom. Consequently,
the coefficients functions of higher dimensional operators computed in the
non-perturbative vacuum may differ from the standard ones computed in the 
perturbative vacuum, even in the standard OPE framework where all power 
corrections are ultimately of IR origin.

The basic diagrammatic quantity which, together with the previous
parameters, determine the power
corrections to a given process is the ``gluon-mass'' dependent
 characteristic function. Its 
discontinuity controls the IR power corrections, which are therefore
related to non-analytic terms in the small gluon mass expansion, whereas
 both the analytic and the non-analytic  terms in this expansion 
 control the UV power corrections. Furthermore, I have checked that
 IR renormalons cancell once the  IR part of the perturbative
calculation  is properly removed; the relevant diagrammatic object
is the ``IR cut-off'' characteristic function.
To establish these
properties for Minkowskian quantities, I used a $Q^2$ analyticity 
approach, whereby each type of power correction is related to the 
time-like discontinuity of the corresponding term in the associated
Euclidean quantity.

I further showed  that, under the assumption of a universal coupling,
 a simple phenomenology of eventual 
``unconventional'' $1/Q^2$ power 
corrections can be developped, focussing on their channel-dependence
 (independently of their physical origin). 
The  evidence for the existence of such terms is presently scarce: they 
have only been detected in a lattice calculation \cite{Marchesini} of the
gluon condensate. A physical picture for their occurence have also been
developped in \cite{AZ}. 

An important issue which deserves further investigation is that of 
renormalization scheme dependence, since the magnitude of the 
exprimentally extracted power corrections depend on the choice of RS.
I have emphasized that the approach of \cite{DW,DMW}, if viewed as the
first (single gluon exchange) term in a yet hypothetical 
``dressed skeleton expansion'' of QCD inherits the essential built-in RS 
independence
of the latter, since the (universal) running coupling 
$\bar{\alpha}_s(k^2)$
 which dresses the 
virtual gluon propagator is supposed to be uniquely 
identified. Furthermore, the potentially dangerous integration over the 
(eventual) Landau 
singularity
at low $k^2$ is avoided in the approach of \cite{DW} through the
introduction of the IR cut-off $\mu_I$, and transmutted into the set of 
non-perturbative
parameters which characterize the low momentum behavior of the coupling. 
The ``infrared
regularized characteristic function'', combined with the
``RS invariant Borel transform'' are the essential tools to perform
the corresponding RS invariant analysis for Minkowskian quantities. 

One might worry about the convergence of the assumed 
``dressed skeleton expansion'', especially if $\bar{\alpha}_s(k^2)$ turns
out to be not particularly small in the infrared region. However, the IR 
magnitude of $\bar{\alpha}_s(k^2)$ may actually  be irrelevant, since 
the
infrared contribution from diagrams with  two (dressed) soft gluon 
exchanges 
(I adopt a 
simple QED analogy for the sake of the argument) is expected to be anyway
power suppressed, whatever the IR magnitude of $\bar{\alpha}_s(k^2)$,
 compare 
to that of the  single (dressed)  soft gluon exchange contribution 
(barring 
eventual problems related to the occurence of the 
``Milan factor''\cite{Mil} in
not inclusive enough Minkowskian
quantities). Some practical questions also remain concerning the optimal
choice of the IR cut-off $\mu_I$, which appears as an effective
additionnal fit parameter in the approach of \cite{DW}.

Finally, this
approach relies on the assumption that the universal QCD 
coupling $\bar{\alpha}_s$ is well defined at the non-perturbative level;
one might then worry whether its perturbative component 
$\bar{\alpha}_s^{PT}$
is also well defined, and   is not itself affected  by the 
IR renormalon ambiguity present in Green's functions.
If this turns out
to be the case, the present framework would become theoretically more 
difficult to justify (even if phenomenologically successful).

\acknowledgments
I have much benefited from discussions with M. Beneke, especially concerning
the contents of section 5.

\appendix

\section{Power corrections in Euclidean quantities}
Given $D(Q^2)$ in eq.(\ref{eq:D}), let us derive the power 
corrections contained in:
\begin{equation}\delta D(Q^2)\equiv \delta D(\Lambda^2/Q^2)=
\int_0^\infty{dk^2\over k^2}\ 
\delta\bar{\alpha}_s(\Lambda^2/k^2)\
\Phi_D(k^2/Q^2)\label{eq:dD}\end{equation}
I assume, for 
$k^2\ll Q^2$, the {\em analytic\/} (if $n$ is integer)  behavior:
\begin{equation}\Phi_D(k^2/Q^2)
\simeq  A\ \left({k^2\over Q^2}\right)^n
+{\cal O}\left[\left({k^2\over Q^2}\right)^{n+1}\right]
\label{eq:D-low}\end{equation}
(this case allows for a 
leading power correction of {\em UV\/} origin in the associated
(section 5)
Minkowskian quantity $R$;
 extension of the present method to deal with 
logarithmic terms in eq.(\ref{eq:D-low})
is straightforward).
On the other hand, for $k^2\gg\Lambda^2$ I assume:
\begin{equation}\delta\bar{\alpha}_s(k^2)\equiv
\delta\bar{\alpha}_s(\Lambda^2/k^2)
\simeq 
\left(b+{c\over\log{k^2\over\Lambda^2}}\right)
\left({\Lambda^2\over
k^2}\right)^n+{\cal O}\left[\left({\Lambda^2\over k^2}\right)^{n+1}\right]
\label{eq:aUV2}\end{equation}
with the {\em same\/} power $n$ (this is the most tricky case, and 
involves
no real loss of generality; the case of inequal powers is dealt
with below). If one then tries to use the expansions 
eq.(\ref{eq:D-low}) or (\ref{eq:aUV2}) under the integral
in eq.(\ref{eq:dD}), one 
encounters UV or IR divergencies. To circumvent this problem, 
it is appropriate to
proceed in a completely symmetrical way with respect
the two relevant scales 
$\Lambda^2$ (the ``small'' scale) and $Q^2$ (the ``large'' scale),
as well as to the two functions $\Phi_D$ and $\delta\bar{\alpha}_s$,
and  split
the integral in eq.(\ref{eq:dD}), at  
$\Lambda^2$ (or, more generally, at $\mu_I^2\geq \Lambda^2$,
to deal with the Landau singularity) and $Q^2$:
\begin{eqnarray}\delta D(Q^2)&=&\int_{0}^{\mu_I^2}
{dk^2\over k^2}\
\delta\bar{\alpha}_s(\Lambda^2/k^2)
\ \Phi_D(k^2/Q^2)
+ \int_{\mu_I^2}^{Q^2}{dk^2\over k^2}\ 
\delta\bar{\alpha}_s(\Lambda^2/k^2)\
\Phi_D(k^2/Q^2)\nonumber\\
&+ & \int_{Q^2}^\infty{dk^2\over k^2}\ 
\delta\bar{\alpha}_s(\Lambda^2/k^2)\
\Phi_D(k^2/Q^2)\label{eq:dD-split} \\
&\equiv&\delta D_{IR}(Q^2)
+\delta D_{UV,(-)}(Q^2)+\delta D_{UV,(+)}(Q^2)\nonumber\end{eqnarray}

The low energy integral $\delta D_{IR}(Q^2)$ is conveniently merged 
with 
the
 corresponding
integral over the perturbative part of the  coupling to yield the term 
$D_{IR}(Q^2)$ in eq.(\ref{eq:D-split})
(the ``infrared'' power
 corrections).
At large $Q^2$ one gets, using eq.(\ref{eq:D-low}):
\begin{equation}D_{IR}(Q^2)\simeq
A\ K(\mu_I^2)\ \left({\Lambda^2\over Q^2}\right)^n
\label{eq:D-ir}\end{equation}
(representing the contribution of a dimension $n$ operator in the OPE),
where:
\begin{equation}K(\mu_I^2)\equiv
\int_{0}^{\mu_I^2}{dk^2\over k^2}\
\bar{\alpha}_s(k^2)\ \left(k^2\over \Lambda^2\right)^n
\label{eq:K-ir}\end{equation}
is a low energy moment of the ``physical'' coupling. 

On the other hand, using eq.(\ref{eq:aUV2}), one gets for the high energy
integral at large $Q^2$:
\begin{equation}\delta D_{UV,(+)}(Q^2)\simeq
\left({\Lambda^2\over Q^2}\right)^n B\left[b+{c\over
\log{Q^2\over\Lambda^2}}+
{\cal O}\left({1\over\log^2{Q^2\over\Lambda^2}}\right)\right]
\label{eq:D-uuv}\end{equation}  
where:
\[B\equiv 
\int_{Q^2}^\infty{dk^2\over k^2}\ \left(Q^2\over k^2\right)^n
\Phi_D(k^2/Q^2)\]

To deal with the ``intermediate range'' integral 
$\delta D_{UV,(-)}(Q^2)$, one uses simultaneously the expansions
eq.(\ref{eq:D-low}) and (\ref{eq:aUV2}), and proceeds by iteration.
 Defining:

\begin{equation}\Phi_D(k^2/Q^2)\equiv A
\left({k^2\over Q^2}\right)^n+\Phi_D^{(n+1)}(k^2/Q^2)
\label{eq:D-low1}\end{equation}
and:
\begin{equation}\delta\bar{\alpha}_s(k^2)\equiv
\left(b+{c\over\log{k^2\over\Lambda^2}}\right)
\left({\Lambda^2\over
k^2}\right)^n+\delta\bar{\alpha}_s^{(n+1)}(k^2)\label{eq:aUV3}
\end{equation}
where $\Phi_D^{(n+1)}(k^2/Q^2)$ is 
${\cal O}\left[\left({k^2\over Q^2}\right)^{n+1}\right]$ 
and $\delta\bar{\alpha}_s^{(n+1)}(k^2)$ is
${\cal O}\left[\left({\Lambda^2\over k^2}\right)^{n+1}\right]$,
 one gets:
\begin{eqnarray}\delta D_{UV,(-)}(Q^2)&=&A\ 
\left({\Lambda^2\over Q^2}\right)^n\ 
\int_{\mu_I^2}^{Q^2}{dk^2\over k^2}\ 
\left(b+{c\over\log{k^2\over\Lambda^2}}\right)\nonumber\\
&+&\left({\Lambda^2\over Q^2}\right)^n\
\int_{\mu_I^2}^{Q^2}{dk^2\over k^2}\
\left({Q^2\over k^2}\right)^n\ \Phi_D^{(n+1)}(k^2/Q^2)
\  \left(b+{c\over\log{k^2\over\Lambda^2}}\right)\nonumber\\
&+&A\ \left({\Lambda^2\over Q^2}\right)^n\
\int_{\mu_I^2}^{Q^2}{dk^2\over k^2}\
\left({k^2\over\Lambda^2}\right)^n\
\delta\bar{\alpha}_s^{(n+1)}(k^2)\nonumber\\
&+& \int_{\mu_I^2}^{Q^2}{dk^2\over k^2}
\ \delta\bar{\alpha}_s^{(n+1)}(k^2)\
\Phi_D^{(n+1)}(k^2/Q^2)\nonumber\end{eqnarray}
Now the last integral yields an 
${\cal O}\left[\left({\Lambda^2\over Q^2}\right)^{n+1}\right]$ 
contribution and can be neglected, whereas, up to ${\cal O}(1/Q^2)$
corrections we have :
\begin{eqnarray}\int_{\mu_I^2}^{Q^2}{dk^2\over k^2}\
\left({Q^2\over k^2}\right)^n\ \Phi_D^{(n+1)}(k^2/Q^2)
\  \left(b+{c\over\log{k^2\over\Lambda^2}}\right)&\simeq&
\int_0^{Q^2}{dk^2\over k^2}\
\left({Q^2\over k^2}\right)^n\ \Phi_D^{(n+1)}(k^2/Q^2)
\  \left(b+{c\over\log{k^2\over\Lambda^2}}\right)\nonumber\\
&=&C\ \left[b+{c\over
\log{Q^2\over\Lambda^2}}+
{\cal O}\left({1\over\log^2{Q^2\over\Lambda^2}}\right)\right]\nonumber
\end{eqnarray}
with:
\[C\equiv 
\int_0^{Q^2}{dk^2\over k^2}\ \left(Q^2\over k^2\right)^n
\Phi_D^{(n+1)}(k^2/Q^2)\]
(where the integral is IR convergent). Furthermore,
up to ${\cal O}(1/Q^2)$
corrections we have:
\[\int_{\mu_I^2}^{Q^2}{dk^2\over k^2}\
\left({k^2\over\Lambda^2}\right)^n\
\delta\bar{\alpha}_s^{(n+1)}(k^2)\simeq
\int_{\mu_I^2}^{\infty}{dk^2\over k^2}\
\left({k^2\over\Lambda^2}\right)^n\
\delta\bar{\alpha}_s^{(n+1)}(k^2)=const\nonumber\]
since the integral is UV convergent. One deduces:
\begin{eqnarray}\delta D_{UV,(-)}(Q^2)
&\simeq& \left({\Lambda^2\over Q^2}\right)^n
\left[A\left(b \log{Q^2\over\Lambda^2}+c  \log\log{Q^2\over\Lambda^2}
\right)
+const+{Cc\over\log{Q^2\over\Lambda^2}}+
{\cal O}\left({1\over\log^2{Q^2\over\Lambda^2}}\right)\right]
\nonumber\\
&&\label{eq:I}\end{eqnarray}
Thus, since:

\begin{equation}\delta D_{UV}(Q^2)\equiv\delta D_{UV,(-)}(Q^2)
+\delta D_{UV,(+)}(Q^2)\label{eq:dDuv-split}
\end{equation}
one ends up with:
\begin{eqnarray}\delta D_{UV}(Q^2)&\simeq&
\left({\Lambda^2\over Q^2}\right)^n
\left[A\left(b\ \log{Q^2\over\Lambda^2}+c\  \log\log{Q^2\over\Lambda^2}
\right)
+const+{(B+C)c\over\log{Q^2\over\Lambda^2}}+
{\cal O}\left({1\over\log^2{Q^2\over\Lambda^2}}\right)\right]
\nonumber\\
& &\label{eq:D-uv1}\end{eqnarray}
Note the log-enhanced terms  arise because I assumed the
unsufficiently suppressed behavior
eq.(\ref{eq:aUV2}) for $\delta\bar{\alpha}_s(k^2)$. Any more damped 
behavior (say with a ${\cal O}(1/\log^2 k^2)$ term, see section 5) 
will only result in
a leading constant term within the brackets in eq.(\ref{eq:D-uv1}).

In the general case where the low $k^2$ behavior of 
$\Phi_D(k^2/Q^2)$
and the high $k^2$ behavior of $\delta\bar{\alpha}_s(k^2)$
 have different 
leading powers, one just  expands either  $\Phi_D(k^2/Q^2)$ 
(at small  $k^2$)
 or $\delta\bar{\alpha}_s(k^2)$ (at large $k^2$)  (depending
 which one has the 
smallest leading power, i.e. is less suppressed)
under the integral in eq.(\ref {eq:dD}), until one is back to the case
of equal powers. In so doing, one never meets any IR or UV divergence
until equal powers are reached. The resulting power terms can then be 
classified as entirely infrared (if $\Phi_D$ is expanded, the standard 
case) or entirely
ultraviolet  (if $\delta\bar{\alpha}_s$ is expanded,
see e.g. eq.(\ref{eq:dD1})).

A similar method allows to derive the large $Q^2$ (or,
equivalently, the small $\mu^2$) expansions of the 
``characteristic functions'' $\dot{{\cal F}}_D(\mu^2/Q^2)$ and 
$\dot{{\cal F}}_R(\mu^2/Q^2)$,
starting from the dispersion relations eq.(\ref{eq:FDdisp}), 
(\ref{eq:Fdisp1}), or (\ref{eq:Fdisp2}) (where $\mu^2$ plays the role
 of
$\Lambda^2$). Note that in the present method, non-analytic
$\log\mu^2$ terms in 
the ``gluon mass'' $\mu^2$ arise from the ``UV'' part of the dispersive
integrals. This fact may cause some confusion, since  these terms
are usually viewed  \cite{BBZ} as the result of {\em IR} divergences.
 This
paradox is clarified by the observation that in the present derivation,
 one chooses $\mu_I^2\simeq \mu^2$, whereas the standard statement is 
correct if one takes $\mu^2\ll \mu_I^2$. This second view point was
 used
 in section 3, where the absence of non-analytic terms at small 
$\mu^2$
 in
$\dot{{\cal F}}_{D,UV}(\mu^2,Q^2)$ and 
$\dot{{\cal F}}_{R,UV}(\mu^2,Q^2)$
 was 
pointed out, despite the presence of $\log Q^2$ terms in their large
$Q^2$
behavior: these functions are the analogues of $\delta D_{UV}$,
 and to derive their large
$Q^2$
behavior  one treats
 $\mu_I^2$ as an ${\cal O}(\mu^2)$ quantity. 
 Note also the non-analytic 
terms may be alternatively viewed as the result of
(gluon mass insensitive) {\em UV} divergences, arising e.g. from taking
the expansion eq.(\ref{eq:D-low}) inside the integral in 
eq.(\ref{eq:FDreg}).

\section{Power corrections in Minkowskian quantities}
When one tries to derive power corrections starting from the 
``Minkowskian'' representation eq.(\ref{eq:R2}), setting:
\begin{equation}\left\{\begin{array}{ccc}\bar{\rho}_s&=&\bar{\rho}_s^{PT}
+\delta\bar{\rho}_s\\
\bar{\alpha}_{eff}&=&\bar{\alpha}_{eff}^{PT}+
\delta\bar{\alpha}_{eff}\end{array}\right.
\label{eq:effsplit}\end{equation}
 one has to take into
account the fact that:
\begin{equation}R_{APT}(Q^2)\equiv
\int_0^\infty{d\mu^2\over\mu^2}
\ \bar{\alpha}_{eff}^{PT}(\mu^2)\ 
\dot{{\cal F}}_R(\mu^2/Q^2)
=R_{PT}(Q^2)+\delta R_{APT}(Q^2)
\label{eq:Rreg}\end{equation}
differs from the Borel sum $R_{PT}$ by power terms $\delta R_{APT}$. They
occur because the ``Minkowskian'' coupling 
$\bar{\alpha}_{eff}^{PT}(\mu^2)$, although an entirely
 perturbative construct, reaches a non-trivial 
IR fixed point at low $\mu^2$ (at the difference of 
$\bar{\alpha}_s^{PT}$).
Similarly, if one defines:
\begin{eqnarray}\bar{\alpha}_s^{APT}(k^2)
&=&-\int_0^\infty{d\mu^2\over\mu^2+k^2}\
\bar{\rho}_s^{PT}(\mu^2)\nonumber\\ &=& k^2\int_0^\infty{d\mu^2
\over(\mu^2+k^2)^2}\
\bar{\alpha}_{eff}^{PT}(\mu^2)\label{eq:areg}\end{eqnarray}
one finds that $\bar{\alpha}_s^{APT}$ 
differs from 
$\bar{\alpha}_s^{PT}$ by power terms $\delta\bar{\alpha}_s^{APT}$:
\begin{equation}\bar{\alpha}_s^{APT}(k^2)=\bar{\alpha}_s^{PT}(k^2)+
\delta\bar{\alpha}_s^{APT}(k^2)\label{eq:dareg}\end{equation}
 which remove the Landau singularity
assumed to be present in $\bar{\alpha}_s^{PT}$.
For instance, in the one-loop case where 
$\bar{\alpha}_s^{PT}(k^2)=1/\beta_0\log(k^2/\Lambda^2)$, one obtains
$\bar{\alpha}_s^{APT}$ by just removing the pole, i.e.:
\begin{equation}\bar{\alpha}_s^{APT}(k^2)={1\over\beta_0\ln(k^2/\Lambda^2)}- 
{1\over\beta_0}{1\over{k^2\over\Lambda^2}-1 }\label{eq:apt-1loop}\end{equation}
In general, one gets, at large $k^2$:
\begin{equation}\delta\bar{\alpha}_s^{APT}(k^2)=\sum_{n=1}^\infty 
(-1)^{n+1}b_n^{APT} 
\left({\Lambda^2\over
k^2}\right)^n\label{eq:daPTas}\end{equation}
where the   constants $b_n^{APT}$ cannot be easily 
calculated for a general perturbative coupling, 
since they depend on all orders \cite{G1} of the perturbative 
beta-function (see eq.(\ref{eq:b_{PT}})
below). $\bar{\alpha}_s^{APT}$ is the ``analytic'' coupling of
 \cite{SS}, 
whose time-like discontinuity coincides\footnote{In this
respect a somewhat artificial element
enters the construction of $\bar{\alpha}_s^{APT}$: beyond one loop, the 
discontinuity  of the (renormalization group improved) perturbative coupling
usually starts in the space-like region. It is therefore arbitrary 
to trunkate this discontinuity at $\mu^2=0$ to enforce a causal coupling:
one could just as well construct an APT coupling whose time-like discontinuity
starts at $\mu^2=c\Lambda^2>0$.}
 with that of the perturbative
coupling.
It is therefore necessary to  split  $\bar{\alpha}_s$  into 
two pieces, each of which satisfies the dispersion 
relation eq.(\ref{eq:disp2}) (at the difference
of the pieces in eq.(\ref{eq:asplit})):
\begin{equation}\bar{\alpha}_s(k^2)=\bar{\alpha}_s^{APT}(k^2)+
\delta\bar{\alpha}_s^{ANP}(k^2)\label{eq:asplit1}\end{equation}
with:
\begin{eqnarray}\delta\bar{\alpha}_s^{ANP}(k^2)&=&-\int_0^\infty{d\mu^2
\over\mu^2+k^2}\ \delta\bar{\rho}_s(\mu^2)\nonumber\\
& =&
 k^2 \int_0^\infty{d\mu^2\over
(\mu^2+k^2)^2}\ \delta\bar{\alpha}_{eff}(\mu^2)\label{eq:daNP}
\end{eqnarray}
and:
\begin{equation}\delta\bar{\alpha}_s(k^2)=\delta\bar{\alpha}_s^{APT}(k^2)+
\delta\bar{\alpha}_s^{ANP}(k^2)\label{eq:dsplit}\end{equation}
($\delta\bar{\alpha}_s$ itself does not  satisfies the 
dispersion relation eq.(\ref{eq:daNP})).
To deal with $\delta\bar{\alpha}_s^{ANP}(k^2)$, I make the simplifying
assumption that $\delta\bar{\alpha}_{eff}(\mu^2)$  is exponentially small
 at large 
$\mu^2$. Then, expanding the kernel under the integral in 
eq.(\ref{eq:daNP}) yields:
\begin{equation}\delta\bar{\alpha}_s^{ANP}(k^2)=\sum_{n=1}^\infty 
(-1)^{n+1} b_n^{ANP} 
\left({\Lambda^2\over
k^2}\right)^n\label{eq:daNPas}\end{equation}
where:
\begin{equation}b_n^{ANP}=\int_0^\infty\ n{d\mu^2\over\mu^2}\ 
\left({\mu^2\over \Lambda^2}\right)^n\
\delta\bar{\alpha}_{eff} (\mu^2)\label{eq:mom}\end{equation}
are integer moments of $\delta\bar{\alpha}_{eff}$. It follows that:
\begin{equation}\delta\bar{\alpha}_s(k^2)=\sum_{n=1}^\infty (-1)^{n+1}b_n 
\left({\Lambda^2\over
k^2}\right)^n\label{eq:daas}\end{equation}
with
\begin{equation}b_n=b_n^{APT}+b_n^{ANP}\label{eq:b-n}\end{equation}
In general, both $\delta\bar{\alpha}_s^{APT}(k^2)$ and 
$\delta\bar{\alpha}_s^{ANP}(k^2)$ are ${\cal O}(1/k^2)$ at large $k^2$, 
but the {\em total\/} coupling modification $\delta\bar{\alpha}_s(k^2)$
itself may decrease faster, if one arranges  the first few ({\em 
or even all}~!)
coefficients $b_n^{APT}$ and $b_n^{ANP}$ to cancell each other.

The proof proceeds by considering separately the contributions 
$ R_{APT}$ and $\delta R_{ANP}$ of 
$\bar{\alpha}_{eff}^{PT}$ and $\delta\bar{\alpha}_{eff}$ respectively 
to:
\begin{equation}R=R_{APT}+\delta R_{ANP}\label{eq:R-split2}
\end{equation}
with:
 \begin{equation}\delta R=\delta R_{APT}+\delta R_{ANP}
\label{eq:dR-split1}\end{equation}

i) Consider first
$R_{APT}(Q^2)$ (eq.(\ref{eq:Rreg})).
 To determine the power terms in $\delta R_{APT}$
(for an Euclidean quantity, they 
would arise
 directly from the contribution of $\delta\bar{\alpha}_s^{APT}$ to 
eq.(\ref{eq:D})), one splits the integral in eq.(\ref{eq:Rreg}) at 
$\mu^2=\Lambda^2$ :
\begin{eqnarray}R_{APT}(Q^2) &=&\int_0^{\Lambda^2}
{d\mu^2\over\mu^2}\ \bar{\alpha}_{eff}^{PT}(\mu^2)\ 
\dot{{\cal F}_R}(\mu^2/Q^2)+
\int_{\Lambda^2}^\infty{d\mu^2\over\mu^2}\ \bar{\alpha}_{eff}^{PT}
(\mu^2)\
\dot{{\cal F}_R}(\mu^2/Q^2)\nonumber\\ 
&\equiv&R_<^{APT}(Q^2)+R_>^{APT}(Q^2)\label{eq:Rregsplit}\end{eqnarray}   
Using  the Borel
representation of $\bar{\alpha}_{eff}^{PT}$ (eq.(\ref{eq:aeffb})), 
$R_>^{APT}$
 can be 
written as:
\begin{equation}R_>^{APT}(Q^2)=R_{PT}(Q^2) - R_<^{PT}(Q^2)\equiv R_>^{PT}(Q^2)
\label{eq:Rapt-pt}\end{equation}
where $R_<^{PT}(Q^2)$ is the Borel sum corresponding to 
$R_<^{APT}(Q^2)$
(this is the piece
 of $R_{PT}$
 which contains the IR renormalons):
\begin{equation}R_<^{PT}(Q^2)\equiv\int_0^\infty dz\ 
\tilde{\alpha}_{eff}(z)\ 
\left[\int_0^{\Lambda^2}{d\mu^2\over \mu^2}\
\dot{{\cal F}_R}(\mu^2/Q^2)\ \exp\left(-z\beta_0\ln{\mu^2\over
\Lambda^2}\right)\right]\label{eq:R_<}\end{equation} 
Thus:

\begin{equation}\delta R_{APT}(Q^2)=R_<^{APT}(Q^2)- R_<^{PT}(Q^2)
\label{eq:dR_{PT}}\end{equation}
 The  power corrections contained in eq.(\ref{eq:dR_{PT}}) are  
obtained 
by expanding $\dot{{\cal F}_R}(\mu^2/Q^2)$
at large $Q^2$ (i.e. low
$\mu^2$)  
 inside the corresponding integrals 
of finite support
$[0,\Lambda^2]$ in eq.(\ref{eq:Rregsplit}) and (\ref{eq:R_<}) 
(note that,
 for $Q^2>\Lambda^2$, 
$\delta R_{APT}(Q^2)$ depends only on the ``low gluon mass piece'' 
${\cal F}_{R,(-)}$ of ${\cal F}_R$).
For instance, an analytic term $n\left({\mu^2\over Q^2}\right)^n$ 
(with $n>0$  integer) in the
low-$\mu^2$ expansion of
$\dot{{\cal F}_R}(\mu^2/Q^2)$ contributes a power correction: 
\begin{equation} b_n^{APT} 
\left({\Lambda^2\over Q^2}\right)^n\label{eq:P1}\end{equation}
with:
\begin{equation}b_n^{APT}=I_n - I_n^{PT}\label{eq:b_{PT}}\end{equation} 
where:
\begin{equation}I_n\equiv 
\int_0^{\Lambda^2}n{d\mu^2\over\mu^2}\left({\mu^2\over
\Lambda^2}\right)^n
\bar{\alpha}_{eff}^{PT}(\mu^2)\label{eq:I_n}\end{equation} 
and:
\begin{eqnarray}I_n^{PT}& \equiv &\int_0^\infty dz\  
\tilde{\alpha}_{eff}(z)\ 
\left[\int_0^{\Lambda^2}n{d\mu^2\over
\mu^2}\
\left({\mu^2\over \Lambda^2}\right)^n\  
\exp\left(-z\beta_0\ln{\mu^2\over \Lambda^2 }\right)\right]\nonumber\\
 &=&\int_0^\infty dz\  
\tilde{\alpha}_{eff}(z)\
\frac{1}{1-\frac{z}{z_n}}\label{eq:I_n^{PT}}
\end{eqnarray}
is the Borel sum corresponding to $I_n$. Applying a similar method to 
the
kernel of the dispersion relation eq.(\ref{eq:areg}) to derive the 
power
corrections in $\delta\bar{\alpha}_s^{APT}$ (eq.(\ref{eq:daPTas}))
shows that the $b_n^{APT}$'s in eq.(\ref{eq:daPTas}) and 
(\ref{eq:b_{PT}})
are actually the same,
which is one way to recover the results for UV power corrections of 
section 5 (at least those that arise from the APT part of the
 coupling).

Similarly, a non-analytic term 
$n\left({\mu^2\over Q^2}\right)^n (c_n \ln {Q^2\over\mu^2} + d_n)$
 in $\dot{{\cal F}_R}$
($n$ integer) contributes a log-enhanced power correction: 
\begin{equation}c_n \left({\Lambda^2\over Q^2}\right)^n 
\left(b_n^{APT} \ln
{Q^2\over\Lambda^2}+\bar{b}_n^{APT}\right) + 
 d_n b_n^{APT} \left({\Lambda^2\over
Q^2}\right)^n\label{eq:P2}\end{equation} 
with:
\begin{equation}\bar{b}_n^{APT}=\bar{I}_n - \bar{I}_n^{PT}
\label{eq:barb_{PT}}\end{equation}
 where:
\begin{equation}\bar{I}_n\equiv\int_0^{\Lambda^2}n{d\mu^2\over\mu^2}
\left({\mu^2\over \Lambda^2}\right)^n\
\ln{\Lambda^2\over\mu^2}\ \bar{\alpha}_{eff}^{PT}(\mu^2)
\label{eq:barI_n}\end{equation} 
and:
\begin{eqnarray}\bar{I}_n^{PT}& \equiv &\int_0^\infty dz\  
\tilde{\alpha}_{eff}(z)\ 
\left[\int_0^{\Lambda^2}n{d\mu^2\over
\mu^2}\
\left({\mu^2\over \Lambda^2}\right)^n\ \ln{\Lambda^2\over\mu^2}\ 
\exp\left(-z\beta_0\ln{\mu^2\over \Lambda^2 }\right)\right]\nonumber\\
 &=&{1\over n}\ \int_0^\infty dz\  
 \tilde{\alpha}_{eff}(z)\
\frac{1}{\left(1-\frac{z}{z_n}\right)^2}\label{eq:bar{I}_n^{PT}}
\end{eqnarray} 
is the Borel sum corresponding to $\bar{I}_n$. 
Note that $\bar{I}_n^{PT}$, hence $\bar{b}_n^{APT}$, are ambiguous, due
 to the presence of
an IR renormalon (a simple pole) at $z=z_n$ in the Borel transform, the 
simple zero in
$\tilde{\alpha}_{eff}(z)$ only partially cancelling the double pole in 
the integrand of
eq.(\ref{eq:bar{I}_n^{PT}}). This is an  example of the relation
 \cite{BBZ,BBB} between non-analytic terms in the
characteristic function and IR renormalons. This relation  can only be
 understood if
$\bar{\alpha}_{eff}^{PT}(\mu^2)$ has a non-trivial IR fixed point: 
otherwise, if one assumes e.g.
$\bar{\alpha}_{eff}^{PT}(\mu^2)$ is given by the one-loop coupling 
(i.e. $\tilde{\alpha}_{eff}(z)
\equiv 1$), one could associate IR renormalons even to analytic terms in
 the low $\mu^2$ expansion
of $\dot{{\cal F}_R}(\mu^2/Q^2)$~! 

\noindent On the other hand, the coefficients
$b_n^{APT}$ of the leading-log parts (and in particular of the analytic
 parts if there are no
accompanying log) are unambiguous for $n$ integer 
(eq.(\ref{eq:I_n^{PT}})), 
which suggests  they should be
associated to short-distances: this interpretation is confirmed by the
results of section 5 and Appendix A .
Note that  {\em all}
 power corrections (both of UV and IR origin) formally
arise (see eq.(\ref{eq:dR_{PT}}) and the remark below) from integration
 over {\em low}
$\mu^2$, and shows  it is  cumbersome   to use the ``Minkowskian'' 
representation eq.(\ref{eq:R2}) to separate long  from short distances, 
at the difference 
 of the ``Euclidean''
 representation eq.(\ref{eq:D}). In particular, $R_>^{APT}(Q^2)$ in
eq.(\ref{eq:Rregsplit}) usually contains ``unorthodox'', OPE unrelated
 UV power 
contributions 
(the exception is the 
case $n\neq integer$), which makes it unconvenient as a definition
of ``regularized'' perturbation theory. It differs from the ``OPE
consistent'' definition $R_{UV}^{PT}(Q^2)$ since it is the
low gluon mass piece (below $\Lambda^2$) of $\dot{{\cal F}_R}$, rather
 then that of its 
discontinuity, which is removed. It may also be misleading, since the 
``unorthodox'' contributions could be removed by similar contributions in
the sum
$R_<^{APT}(Q^2)+\delta R_{ANP}(Q^2)$ (see below).

ii) Consider next the contribution of $\delta\bar{\alpha}_{eff}$:
\begin{equation}\delta R_{ANP}(Q^2)\equiv 
\int_0^\infty{d\mu^2\over\mu^2}\ 
\delta\bar{\alpha}_{eff}(\mu^2)\ 
\dot{{\cal F}}_R(\mu^2/Q^2)
\label{eq:dRNP}\end{equation}
If one makes again the 
simplifying assumption
that $\delta\bar{\alpha}_{eff}(\mu^2)$ is exponentially suppressed
at large $\mu^2$, the corresponding power corrections
are obtained  by taking the low $\mu^2$ expansion of
$\dot{{\cal F}_R}(\mu^2/Q^2)$ inside the integral in 
eq.(\ref{eq:dRNP}). For instance,
  a non-analytic term
$n\left({\mu^2\over Q^2}\right)^n (c_n \ln{Q^2\over\mu^2} + d_n)$
 in $\dot{{\cal F}_R}$ 
($n$ integer) 
contributes a log-enhanced
power correction:
\begin{equation}c_n \left({\Lambda^2\over Q^2}\right)^n \left(b_n^{ANP} 
\ln{Q^2\over\Lambda^2}+\bar{b}_n^{ANP}\right) +  d_n b_n^{ANP} 
\left({\Lambda^2\over Q^2}\right)^n\label{eq:PNP}\end{equation} 
where $b_n^{ANP}$ is given in eq.(\ref{eq:mom}), and:
\begin{equation}\bar{b}_n^{ANP}=\int_0^\infty n{d\mu^2\over\mu^2}
\left({\mu^2\over \Lambda^2}\right)^n\
\ln{\Lambda^2\over\mu^2} \ \delta\bar{\alpha}_{eff}(\mu^2)
\label{eq:bar{b}_n^{NP}}\end{equation}
 Note that eq.(\ref{eq:PNP}) has
{\em exactly} the same structure as the corresponding contribution to
$\delta R_{APT}(Q^2)$ (eq.(\ref{eq:P2})) with the substitutions 
$b_n^{APT}\rightarrow b_n^{ANP}$ and
$\bar{b}_n^{APT}\rightarrow \bar{b}_n^{ANP}$! Again, the leading log
 terms terms with a
coefficient $b_n^{ANP}$ (an {\em analytic, integer} moment) should be 
associated to short
distances, while the sub-leading log terms, with a coefficient 
$\bar{b}_n^{ANP}$ (a {\em non-analytic} moment), are partly long distance.

\noindent\underline{Application to the causal perturbative coupling}:
this is the case where $\bar{\alpha}_s^{PT}\equiv\bar{\alpha}_s^{APT}$,
and $\delta\bar{\alpha}_s^{APT}\equiv 0$. Then the $b_n^{APT}$'s must all
vanish, which implies that the power corrections in $R_{APT}(Q^2)$ are
``OPE compatible'', since only non-analytic terms in $\dot{{\cal F}_R}$
can contribute.

\noindent Furthermore, for a general $\bar{\alpha}_s^{PT}$ (not necessarily 
causal) one can show that power corrections in $R_{UV}^{APT}(Q^2)$ 
involve only the $b_n^{APT}$'s, and therefore all vanish when
specialized to a causal
coupling. To prove the former statement, I note that a large $Q^2$
contribution to $\dot{{\cal F}}_{R,UV}(\mu^2,Q^2)$ of the form
$n\left({\mu^2\over Q^2}\right)^n\left (c_n \ln{Q^2\over\Lambda^2} 
+  d_n(\mu^2/\Lambda^2)\right)$ (I take for simplicity the IR cut-off 
$\mu_I$
equal to $\Lambda$) contributes a power correction:
\begin{equation} \left({\Lambda^2\over Q^2}\right)^n \left(c_n b_n^{APT} 
\ln{Q^2\over\Lambda^2}+J_n-J_n^{PT}\right) \label{eq:PPTUV}\end{equation} 
where:
\begin{equation}J_n\equiv \int_0^{\Lambda^2}n{d\mu^2\over\mu^2}
\left({\mu^2\over \Lambda^2}\right)^n\
d_n\left({\mu^2\over\Lambda^2}\right)\ \bar{\alpha}_{eff}^{PT}(\mu^2)
\label{eq:J_n}\end{equation}
and $J_n^{PT}$ is the corresponding Borel sum. But $d_n(\mu^2/\Lambda^2)$
is {\em analytic} at small $\mu^2$ (see the comments at the end of 
Appendix A), and consequently, expanding  $d_n(\mu^2/\Lambda^2)$ in 
powers of $\mu^2$ 
 under the integral in
eq.(\ref{eq:J_n}) and its analogue in $J_n^{PT}$ (it is easy to 
check that the radius of convergence of the series is 
$\mu^2/\Lambda^2=1$) yields only terms
 proportionnal to the {\em integer} moments $b_p^{APT}$ .

\section{Expressing Minkowskian power corrections in term of 
$\delta\bar{\alpha}_s$}
Let us give a proof of eq.(\ref{eq:dR}) alternative to that of 
\cite{G1}.
The proof proceeds again by considering separately the contributions 
$R_{APT}$ and $\delta R_{ANP}$ of 
$\bar{\alpha}_{eff}^{PT}$ and $\delta\bar{\alpha}_{eff}$.

i) Consider first
$R_{APT}(Q^2)$ (eq.(\ref{eq:Rreg})).
 An explicit expression for $\delta R_{APT}$ 
has been
obtained in the one-loop case in \cite{BBB}. The following can be seen as
 an extension  of their result. Splitting the integral in 
eq.(\ref{eq:Rreg}) at $Q^2$, it is clear the piece above $Q^2$ 
contributes only to the Borel sum. 
It is therefore sufficient to consider:
\begin{equation}R_{(-)}^{APT}(Q^2)\equiv\int_0^{Q^2}
{d\mu^2\over\mu^2}
\ \bar{\alpha}_{eff}^{PT}(\mu^2)\ 
\dot{{\cal F}}_{R,(-)}(\mu^2/Q^2)\label{eq:Rreg-}\end{equation}
which shares the same power terms as $R_{APT}$. 
The expression for
 $\delta R_{APT}$ depends on the form of the dispersion relation
 satisfied
by ${\cal F}_{R,(-)}$. 

Assume for simplicity one subtraction at $\mu^2=0$.
Then:
\begin{equation}\dot{{\cal F}}_{R,(-)}(\mu^2/Q^2)=
\mu^2\int_0^\infty{dk^2\over(k^2+\mu^2)^2}\ 
\Phi_R(k^2/Q^2)\label{eq:Fdisp1}\end{equation}
Substituting eq.(\ref{eq:Fdisp1}) into eq.(\ref{eq:Rreg-}) yields:
\begin{equation}R_{(-)}^{APT}(Q^2)=\int_{0}^\infty{dk^2\over k^2}\ 
\bar{\alpha}_{s,(-)}^{APT}(k^2,Q^2)\ 
\Phi_R(k^2/Q^2)
\label{eq:Rreg-1}
\end{equation}
with:
\begin{equation}\bar{\alpha}_{s,(-)}^{APT}(k^2,Q^2)
\equiv k^2\int_0^{Q^2}{d\mu^2
\over(\mu^2+k^2)^2}\ \bar{\alpha}_{eff}^{PT}(\mu^2)\label{eq:a-}
\end{equation}
It is clear however that the power terms in $\bar{\alpha}_{s,(-)}^{APT}$ 
are the
 same as those in $\bar{\alpha}_s^{APT}$, since again the integration 
range
 above
$Q^2$ in the dispersion relation eq.(\ref{eq:areg}) contributes only to 
the Borel sum. Thus:
\[\bar{\alpha}_{s,(-)}^{APT}(k^2,Q^2)=\bar{\alpha}_{s,(-)}^{PT}(k^2,Q^2)+
\delta\bar{\alpha}_s^{APT}(k^2)\]
(where $\bar{\alpha}_{s,(-)}^{PT}$ is defined as a Borel sum)
and:
\[R_{(-)}^{APT}(Q^2)=R_{(-)}^{PT}(Q^2)+\delta R_{APT}(Q^2)\]
 with:
\[R_{(-)}^{PT}(Q^2)=\int_{0}^\infty{dk^2\over k^2}
\  \bar{\alpha}_{s,(-)}^{PT}
(k^2,Q^2)\ \Phi_R(k^2/Q^2)\]
and:
\begin{equation}\delta R_{APT}(Q^2)=\int_0^\infty{dk^2\over k^2}
\ \delta\bar{\alpha}_s^{APT}(k^2)\
\Phi_R(k^2/Q^2) \label{eq:dRPT}\end{equation}

Assume next two subtractions at $\mu^2=0$. Then:
\begin{equation}\dot{{\cal F}}_{R,(-)}(\mu^2/Q^2)=
a_0{\mu^2
\over Q^2}+
\mu^2\int_0^\infty dk^2\left[{1\over(k^2+\mu^2)^2}-{1\over k^4}\right]
\ \Phi_R(k^2/Q^2)\label{eq:Fdisp2}\end{equation}
where $a_0$ is a subtraction constant, and 
$\Phi_R(k^2/Q^2)$ is assumed to be 
${\cal O}(k^4/Q^4)$ at small $k^2$, in order to have no IR divergence
in the dispersive integral. Substituting 
eq.(\ref{eq:Fdisp2}) into eq.(\ref{eq:Rreg-}) yields:
\begin{equation}R_{(-)}^{APT}(Q^2)=a_0\ R_1^{APT}(Q^2)+
\int_{0}^\infty{dk^2\over k^2}\  \left[
\bar{\alpha}_{s,(-)}^{APT}(k^2,Q^2)-{Q^2\over k^2}\ R_1^{APT}(Q^2)
\right]\ \Phi_R(k^2/Q^2)
\label{eq:Rreg-2}\end{equation}
where:
\begin{equation}R_1^{APT}(Q^2)\equiv \int_0^{Q^2}{d\mu^2\over\mu^2}
{\mu^2\over Q^2}
\bar{\alpha}_{eff}^{PT}(\mu^2)\label{eq:R1apt}\end{equation}
 Again, $R_1^{APT}(Q^2)$ differs from its Borel sum
$R_1^{PT}(Q^2)$ by a power correction $b_1^{APT}{\Lambda^2\over Q^2}$,
 and one
 can easily show \cite{G1} with the method of Appendix B 
 that $b_1^{APT}$ is the same coefficient which 
appears in
 the 
leading term in eq.(\ref{eq:daPTas}). The power corrections in 
eq.(\ref{eq:Rreg-2}) are therefore obtained by substituting $R_1^{APT}$
 and
$\bar{\alpha}_{s,(-)}^{APT}$ by they respective power corrections pieces 
$b_1^{APT}{\Lambda^2\over Q^2}$ and $\delta\bar{\alpha}_s^{APT}(k^2)$,
which yields:
\begin{equation}\delta R_{APT}(Q^2)=a_0\ b_1^{APT}{\Lambda^2\over
 Q^2}+
\int_0^\infty{dk^2\over k^2}\ 
\left[\delta\bar{\alpha}_s^{APT}(k^2)-b_1^{APT}{\Lambda^2\over k^2}\right
]\ \Phi_R(k^2/Q^2) \label{eq:dRPT1}\end{equation}
where the integrand is properly subtracted to insure convergence
at large $k^2$
(similar expressions may be obtained if the subtractions are performed 
away from $\mu^2=0$).

ii) Consider next $\delta R_{ANP}(Q^2)$  (eq.(\ref{eq:dRNP})).
If one makes again the 
simplifying assumption
that $\delta\bar{\alpha}_{eff}(\mu^2)$ is exponentially suppressed
at large $\mu^2$, one gets:
\begin{equation}\delta R_{ANP}(Q^2)\simeq \int_0^\infty{d\mu^2
\over\mu^2}
\ \delta\bar{\alpha}_{eff}(\mu^2)\ 
\dot{{\cal F}}_{R,(-)}(\mu^2/Q^2) \label{eq:dR-NP}\end{equation}
up to exponentially small corrections at large $Q^2$. It is then 
straightforward to express the right-hand side of eq.(\ref{eq:dR-NP})
in term of $\delta\bar{\alpha}_s^{ANP}$. Assuming for instance the  
dispersion relation eq.(\ref{eq:Fdisp2}), and using
the dispersion relation eq.(\ref{eq:daNP}), one gets:
\begin{equation}\delta R_{ANP}(Q^2)\simeq a_0\ b_1^{ANP}
{\Lambda^2\over Q^2}+
\int_0^\infty{dk^2\over k^2}\ 
\left[\delta\bar{\alpha}_s^{ANP}(k^2)-b_1^{ANP}{\Lambda^2\over k^2}\right
]\ \Phi_R(k^2/Q^2) \label{eq:dRNP1}\end{equation}
with $b_1^{ANP}$ given in eq.(\ref{eq:mom}). Eq.(\ref{eq:dRNP1}) has 
the same
 form as eq.(\ref{eq:dRPT1}). Adding the two yields the final result
for $\delta R$:
\begin{equation}\delta R(Q^2)\simeq a_0\ b_1{\Lambda^2\over Q^2}+
\int_0^\infty{dk^2\over k^2}\ 
\left[\delta\bar{\alpha}_s(k^2)-b_1{\Lambda^2\over k^2}\right
]\ \Phi_R(k^2/Q^2) \label{eq:dR2}\end{equation}
where $b_1=b_1^{APT}+b_1^{ANP}$,
which is independent of the split in eq.(\ref{eq:asplit1}), and is 
correct at large $Q^2$ up to exponentially small corrections
 within the present assumptions. If one further assumes that
$\delta\bar{\alpha}_s(k^2)$ decreases faster then $1/k^2$, then $b_1=0$,
 and one recovers eq.(\ref{eq:dR}) (faster decrease may be necessary 
to make the right hand side of eq.(\ref{eq:dR}) ultraviolet convergent,
 depending on the number of assumed subtractions in the dispersion
relation for ${\cal F}_{R,(-)}$). Note also the
results of this section  are also valid in the more general case where 
${\cal F}_{R,(-)}(\mu^2/Q^2)$ can be written as the sum of a function 
which
 satisfies a dispersion
relation (hence has no complex singularities) and a function analytic 
around the origin - a
generalized ``subtraction term'' (but which may have complex 
singularities at finite distance
from the origin, i.e. for large enough $\mu^2/Q^2$).


\end{document}